\documentclass[sigconf]{acmart}

\AtBeginDocument{%
  }

\usepackage{xcolor}
\usepackage{colortbl}
\usepackage{graphicx}
\usepackage{booktabs}
\usepackage{array}
\usepackage{multirow}
\usepackage{makecell}
\usepackage{tikz}
\usepackage{adjustbox}
\usepackage{xspace}

\newcommand{\siteurl}{%
  \href{https://agenticva.github.io/AgenticVA/}{%
    \texttt{agenticva.\allowbreak github.\allowbreak io/\allowbreak AgenticVA/}}%
}

\makeatletter
\renewcommand\showeprint[2][arxiv]{%
  \def\@tempa{#1}%
  \ifx\@tempa\@empty\def\@tempa{arxiv}\fi
  \def\@tempb{arxiv}%
  \ifx\@tempa\@tempb\relax
     arXiv:\href{https://arxiv.org/abs/#2}{\nolinkurl{#2}}%
  \else
     \def\@tempb{arXiv}%
     \ifx\@tempa\@tempb\relax
       arXiv:\href{https://arxiv.org/abs/#2}{\nolinkurl{#2}}%
     \else
       #1:\nolinkurl{#2}%
    \fi
  \fi}
\makeatother

\newcommand{\rot}[1]{\rotatebox[origin=c]{90}{#1}}
\newcommand{\nistart}[1]{{\noindent\textbf{#1}}}
\definecolor{red}{RGB}{184,84,80}
\definecolor{green}{RGB}{0,128,0}
\definecolor{blue}{RGB}{33,89,202}
\definecolor{orange}{RGB}{215,155,0}
\definecolor{purple}{RGB}{150,115,166}
\definecolor{cyan}{RGB}{14,128,136}
\definecolor{dawn}{RGB}{161,156,145}
\definecolor{sand}{RGB}{204,103,1}

\newcommand{\red}[1]{\textcolor{red}{#1}}
\newcommand{\green}[1]{\textcolor{green}{#1}}
\newcommand{\blue}[1]{\textcolor{blue}{#1}}
\newcommand{\orange}[1]{\textcolor{orange}{#1}}
\newcommand{\purple}[1]{\textcolor{purple}{#1}}
\newcommand{\cyan}[1]{\textcolor{cyan}{#1}}
\newcommand{\dawn}[1]{\textcolor{dawn}{#1}}
\newcommand{\sand}[1]{\textcolor{sand}{#1}}

\newcommand{\boldcheck}{\tikz{\fill (0,0) circle(0.7ex);}}

\newcommand{\stab}{\vspace{1.2ex}\noindent}
\newcommand{\stitle}[1]{\stab\noindent{\bf #1}}

\newcommand{\planner}{\purple{\textbf{\textsc{Planner}}}\xspace}
\newcommand{\creator}{\sand{\textbf{\textsc{Creator}}}\xspace}
\newcommand{\reviewer}{\cyan{\textbf{\textsc{Reviewer}}}\xspace}
\newcommand{\contextmanager}{\textbf{\dawn{\textsc{ContextManager}}}\xspace}

\newcommand{\dataprocessing}{\blue{\textbf{\textsc{DataProcessing}}}\xspace}
\newcommand{\visualmapping}{\orange{\textbf{\textsc{VisualMapping}}}\xspace}
\newcommand{\viewtransformation}{\red{\textbf{\textsc{ViewTransformation}}}\xspace}
\newcommand{\presentation}{\textbf{\green{\textsc{Presentation}}}\xspace}

\newcommand*\colorcircled[3]{\tikz[baseline=(char.base)]{
            \node[shape=circle,fill=#1,text=#2,inner sep=0.5pt,font=\footnotesize\bfseries] (char) {#3};}}
\newcommand{\levelone}{\colorcircled{orange}{white}{1}}
\newcommand{\leveltwo}{\colorcircled{red}{white}{2}}
\newcommand{\levelthree}{\colorcircled{cyan}{white}{3}}
\newcommand{\levelfour}{\colorcircled{purple}{white}{4}}

\newcommand{\implicationbox}[2]{%
  \par\vskip 1ex
  \noindent\fcolorbox{cyan!50}{cyan!5}{%
    \parbox{\dimexpr\linewidth-2\fboxsep-2\fboxrule\relax}{\textbf{Implication #1.} #2}%
  }%
  \par\vskip 1ex
}

\begin{document}

\title{Exploring Agentic Visual Analytics: \\ A Co-Evolutionary Framework of Roles and Workflows}

\author{Tianqi Luo}
\email{tluo553@connect.hkust-gz.edu.cn}
\affiliation{%
  \institution{The Hong Kong University of Science and Technology (Guangzhou)}
  \city{Guangzhou}
  \country{China}
}

\author{Leixian Shen}
\email{lshenaj@connect.ust.hk}
\affiliation{%
  \institution{Microsoft Research}
  \city{Redmond}
  \state{WA}
  \country{USA}
}
\affiliation{%
  \institution{The Hong Kong University of Science and Technology}
  \city{Hong Kong SAR}
  \country{China}
}

\author{Mingyue Yuan}
\email{mingyue.yuan@unsw.edu.au}
\affiliation{%
  \institution{University of New South Wales}
  \city{Sydney}
  \country{Australia}
}

\author{Yuyu Luo}
\email{yuyuluo@hkust-gz.edu.cn}
\affiliation{%
  \institution{The Hong Kong University of Science and Technology (Guangzhou)}
  \city{Guangzhou}
  \country{China}
}

\renewcommand{\shortauthors}{Luo et al.}

\begin{abstract}%
% CHI 2027 requires abstracts of 150 words or fewer; current version: ~146 words.
 In agentic visual analytics (VA), large language model (LLM)-driven agents autonomously plan, execute, evaluate, and iterate across the VA pipeline.
 Agentic delegation is changing human--data interaction, yet its rapid growth has outpaced understanding of the design landscape. We ask \textit{how agent responsibilities are allocated across VA roles and workflow stages}, and \textit{how this allocation reshapes human responsibilities and burdens}.
 We survey 65 agentic VA systems and introduce a \textit{co-evolutionary} framework that maps rising agent autonomy against the redistribution of human operational, verification, and governance responsibilities.
 Within this framework, a role-workflow taxonomy aligns four agentic roles (\planner, \creator, \reviewer, \contextmanager) with established VA pipeline stages.
 We distill trade-offs across autonomy levels, roles, and workflow stages into situated, evidence-grounded design implications and future research directions.
 An interactive browser of the framework, corpus, and guidelines is available at \siteurl.
\end{abstract}


\begin{teaserfigure}
  \includegraphics[width=\textwidth]{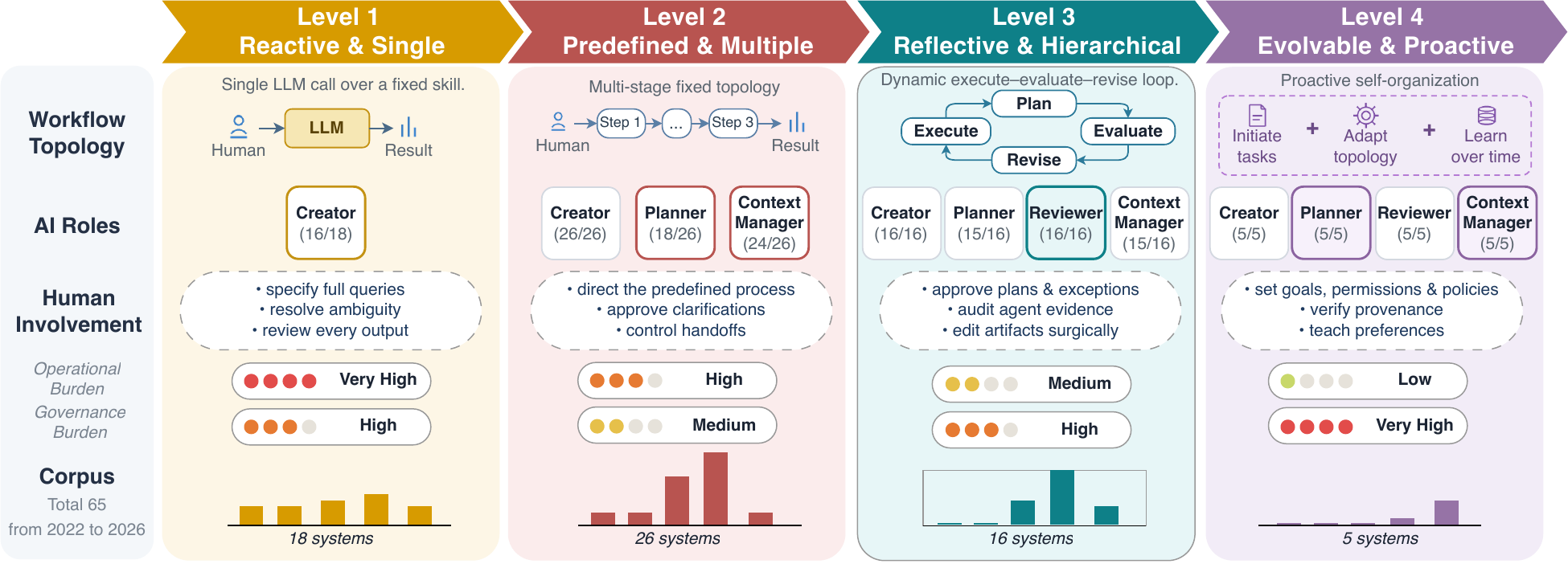}
  % \vspace{-0em}
  \caption{%
  	The Co-Evolutionary Framework of agentic visual analytics. As autonomy rises from Level~1 (Reactive) to Level~4 (Proactive--Evolvable), role allocation accumulates: a closed-loop runtime \reviewer{} that drives revision marks the Level~3 Reflective--Adaptive gate, while at Level~4 the \planner{} can self-initiate and the \contextmanager{} can evolve into a skill or knowledge base. Human duties shift from operating toward governing, and two burdens diverge on a shared scale: operational burden falls monotonically, whereas verification and governance burden falls and then rises. Counts are drawn from our 65-system corpus.
  }
  \Description{A grid with four columns for autonomy Levels 1 through 4 and seven rows for workflow topology, cumulative AI roles, human duties, control mechanisms, operational burden, governance burden, and corpus footprint. Workflow topology grows from a single LLM call over a fixed skill, through a predefined multi-stage workflow, to a dynamic execute-evaluate-revise loop at Level 3, and proactive self-organization at Level 4. AI roles accumulate from Creator, adding Planner and Context Manager, and then the runtime Reviewer as the Level 3 gate; at Level 4, Planner self-initiates and Context Manager evolves from session state into a skill or knowledge base. Human duties move from specifying and reviewing each operation toward setting goals, permissions, and policies, verifying provenance, and teaching preferences. Control expands from natural language to multi-turn interfaces, plan editing, behavioral sensing, and policy configuration. Operational burden falls from very high to low, while governance burden falls from high to medium and then rises to very high. The corpus footprint is 18, 26, 16, and 5 systems across Levels 1 through 4.}
  \label{fig:teaser}
\end{teaserfigure}

\maketitle

%!TEX root = ../main.tex
\section{Introduction}
\label{sec:intro}

Visual analytics (VA) has traditionally required people to operate analytical tools directly: they prepare data, select visual encodings, manipulate views, and assemble findings for communication~\cite{10.1145/1809400.1809403,8740868}. Natural-language interfaces lower some of this interaction cost, but most early systems remain reactive---they translate a request into an operation or visualization and return control to the user~\cite{hutchinson2024LLMAssistedVisualAnalytics,narechania2021NL4DVToolkitGenerating,luo2022NaturalLanguageVisualization}. LLM-driven agents are changing this arrangement. Recent systems can decompose goals, invoke tools, inspect intermediate outputs, revise failures, and coordinate work across multiple analytical stages~\cite{yao2023ReActSynergizingReasoning,dibia2023LIDAToolAutomatic,chen2025codaagenticsystemscollaborative}. The interaction problem is therefore no longer only how people issue commands to analytical systems, but how analytical work and responsibility are divided between people and increasingly autonomous agents.

Prior surveys explain how AI supports visualization~\cite{wu2022AI4VISSurveyArtificial,wang2022SurveyML4VISApplying}, how natural language can mediate visual analysis~\cite{shen2023NaturalLanguageInterfaces,zhang2024NaturalLanguageInterfaces}, and how LLM-based agents plan and coordinate~\cite{wang2024SurveyLargeLanguage,guo2024LargeLanguageModel,zhu2026surveydataagentsemerging}. Together, this literature provides increasingly detailed accounts of what such systems can do. It offers less support, however, for understanding how increasing autonomy changes the \emph{division of analytical labor}: which responsibilities move to agents, where those responsibilities enter the VA workflow, and which responsibilities remain or newly emerge for people. This gap matters because delegating execution does not delegate accountability. Errors in data preparation, visual mapping, or generated narratives can propagate across a workflow while becoming harder for users to observe.

Our \textbf{Co-Evolutionary Framework} (Fig.~\ref{fig:teaser}) characterizes this transition through four autonomy regimes: Level~1 \emph{Reactive}, Level~2 \emph{Predefined Workflow}, Level~3 \emph{Reflective--Adaptive}, and Level~4 \emph{Proactive--Evolvable}. These levels do not simply count automated steps. They mark qualitative changes in who can decide, evaluate, revise, initiate, and govern analytical work. As autonomy rises, agent responsibilities accumulate and reorganize, while human responsibilities shift from specifying and executing individual operations toward inspecting plans and evidence, setting permissions, checking provenance, and governing persistent knowledge. Our central finding is that \textbf{increasing autonomy redistributes human work rather than removing it}: operational burden falls as execution is delegated, whereas verification and governance burden initially falls but rises again when users must oversee reflective, proactive, and persistent agents.

\begin{figure}[t]
  \centering
  \includegraphics[width=\columnwidth]{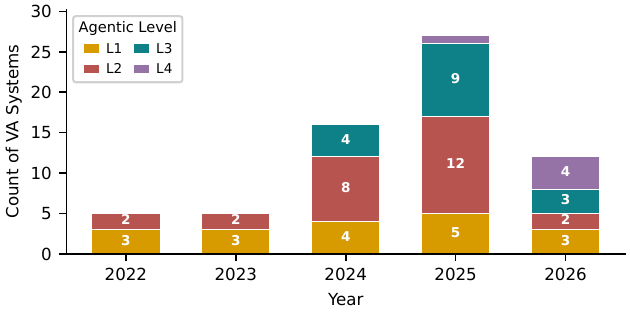}
  \vspace{-1em}
  \caption{Distribution of the 65 surveyed systems by publication year and autonomy level. Data collection concluded in August 2026.}
  \Description{A stacked bar chart showing surveyed systems by publication year and four autonomy levels. Publication counts rise sharply in 2025 and 2026, and Levels 3 and 4 are concentrated in recent years. Data collection concluded in August 2026.}
  \label{fig:timeline}
  \vspace{-2em}
\end{figure}

This transition is already visible in the literature. We identify 65 primary systems published between 2022 and August 2026. As Fig.~\ref{fig:timeline} shows, publication activity increases sharply in 2025--2026, when nearly all Level~3 and Level~4 systems in our corpus appear. The trajectory is not merely toward better reactive generation: recent systems increasingly incorporate dynamic planning, output-driven reflection, proactive intervention, and persistent adaptation. This empirical shift motivates a systematic examination of how higher autonomy changes both system architecture and human involvement.

\begin{figure*}[t]
  \centering
  \includegraphics[width=.8\textwidth]{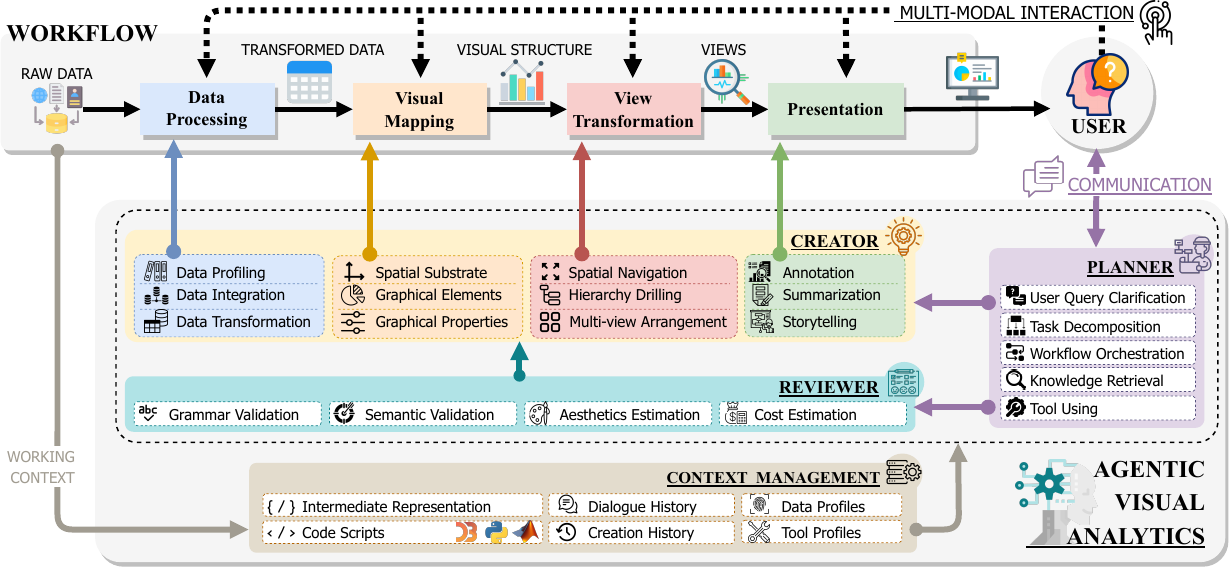}
  \caption{The analytical anatomy of agentic visual analytics. We code system functions by four cross-cutting roles (\planner, \creator, \reviewer, and \contextmanager) and locate their work within four VA workflow stages (\dataprocessing, \visualmapping, \viewtransformation, and \presentation). Autonomy level is the macro organizing axis; Role~$\times$~Workflow provides the micro-level lens for explaining differences between levels.}
  \Description{A diagram of an agentic visual analytics workflow connecting a user, raw data, working context, and views. Planner, Creator, Reviewer, and Context Manager functions operate across data processing, visual mapping, view transformation, and presentation, with multimodal interaction linking users to the workflow.}
  \label{fig:overview}
  \vspace{-1em}
\end{figure*}

To explain why these autonomy regimes differ, we decompose each system along two orthogonal dimensions (Fig.~\ref{fig:overview}). The first asks \emph{what function performs the work}: the \planner{} clarifies and decomposes goals, the \creator{} produces analytical artifacts, the \reviewer{} evaluates artifacts and can trigger revision, and the \contextmanager{} maintains the state and knowledge needed across actions. The second asks \emph{where the work occurs}: \dataprocessing, \visualmapping, \viewtransformation, or \presentation. Autonomy is thus our macro organizing axis, while Role~$\times$~Workflow is the micro analytical lens through which we examine its internal structure.

We organize the review around two research questions:
\begin{description}
  \item[\textbf{RQ1.}] \emph{As system autonomy increases, how are agentic responsibilities allocated across functional roles and VA workflow stages?}
  \item[\textbf{RQ2.}] \emph{How does this allocation reshape human responsibilities, interaction and control mechanisms, and operational versus verification and governance burdens?}
\end{description}
RQ1 examines changes on the agent side; RQ2 examines the corresponding redistribution of human work. Together, they operationalize our co-evolutionary account of agentic VA.

This paper makes three contributions:
\begin{itemize}
  \item \textbf{An empirical foundation:} a systematic corpus of 65 agentic VA systems, with an operational coding scheme for autonomy, functional roles, workflow stages, and human interaction and control.
  \item \textbf{A co-evolutionary framework:} an integrated account linking autonomy progression to Role~$\times$~Workflow reconfiguration and the redistribution of human analytical responsibilities.
  \item \textbf{A human-centered synthesis:} evidence showing how human work shifts from operation toward verification and governance, together with situated implications for designing controllable, inspectable, and accountable agentic VA systems.
\end{itemize}
% \input{src/2.related}
%!TEX root = ../main.tex
\section{Methodology}
\label{sec:method}

\subsection{Systematic Corpus Construction}
\label{sec:corpus}

% We conducted a systematic literature review guided by PRISMA~\cite{page2021PRISMA2020}. Our search covered research published from 2022 to August 2026, a period beginning with the broad availability of instruction-tuned LLMs capable of tool use and multi-step reasoning. We searched major venues across visualization (e.g., IEEE VIS, TVCG, and EuroVis), human--computer interaction (e.g., CHI and UIST), data management (e.g., VLDB and SIGMOD), and AI and NLP (e.g., ACL, NeurIPS, and ICLR). Keyword-based searches were supplemented with backward and forward citation tracking.

We constructed the corpus through a structured literature search covering work published from 2022 to August 2026, a period beginning with the broad availability of instruction-tuned LLMs capable of tool use and multi-step reasoning. We searched major venues across visualization (e.g., IEEE VIS, TVCG, and EuroVis), human–computer interaction (e.g., CHI and UIST), data management (e.g., VLDB and SIGMOD), and AI and NLP (e.g., ACL, NeurIPS, and ICLR), using combinations of terms related to large language models, agents, visualization, and visual analytics. We supplemented keyword-based retrieval with backward and forward citation tracking of included papers and prior surveys.

We included work that (1) presents a system that directly supports visual data analysis and (2) incorporates at least one LLM-driven function for planning, creation, review, or context management. We excluded systems that use an LLM only for isolated summarization without generating, modifying, or interacting with visual analytical artifacts. We also excluded earlier rule-based NLIs, which are well documented by prior surveys~\cite{shen2023NaturalLanguageInterfaces,zhang2024NaturalLanguageInterfaces}, except when they provide a necessary baseline. Benchmarks and empirical studies were retained as supporting evidence but were not coded as primary systems.

The final corpus contains \textbf{65 primary systems} and \textbf{14 separately tracked benchmarks and empirical studies}. Table~\ref{tab:survey_papers_final} summarizes the primary systems; the companion website provides the searchable corpus and coding details.

%!TEX root = ../main.tex

\begin{table*}[tp]
    \centering
    \caption{Overview of agentic visual analytics systems (2022–2026) by autonomy level, agentic roles, workflow stages, and interaction. Benchmarks and empirical studies are accounted separately (see supplement).}
    \label{tab:survey_papers_final}
    \vspace{-6pt}
    
    % The table is narrower now, so \small or \footnotesize is a good size.
    \scriptsize
    
    % Updated tabular environment: added Category column before Agents
    % \begin{tabular}{|c|p{3cm}|c|*{4}{c}|*{12}{c}|*{2}{c}|*{3}{c}|}
    \renewcommand\arraystretch{0.72}
\setlength{\tabcolsep}{1.45mm}{
    \begin{tabular}{|c|p{3.3cm}|>{\centering\arraybackslash}p{0.5cm}|*{4}{>{\centering\arraybackslash}p{0.15cm}}|*{12}{>{\centering\arraybackslash}p{0.15cm}}|*{2}{>{\centering\arraybackslash}p{0.3cm}}|}
        % \hline
        \cline{3-21}
        
        % --- TOP HEADER ROW ---
        \multicolumn{2}{c|}{} &
        \multicolumn{1}{c|}{} &
        \multicolumn{4}{c|}{Agents} &
        \multicolumn{12}{c|}{Visual Analytic Workflow} &
        \multicolumn{2}{c|}{Interact} \\
        % \hline
        \cline{3-21}

        % --- SECOND HEADER ROW ---
        \multicolumn{2}{c|}{} &
        \rotatebox[origin=l]{90}{Level} &
        \rotatebox[origin=l]{90}{\purple{Planner}} &
        \rotatebox[origin=l]{90}{\sand{Creator}} &
        \rotatebox[origin=l]{90}{\cyan{Reviewer}} &
        \rotatebox[origin=l]{90}{\dawn{Context Manager}} &
        \rotatebox[origin=l]{90}{\blue{Data Profiling}} &
        \rotatebox[origin=l]{90}{\blue{Data Integration}} &
        \rotatebox[origin=l]{90}{\blue{Data Transformation}} &
        \rotatebox[origin=l]{90}{\orange{Spatial Substrate}} &
        \rotatebox[origin=l]{90}{\orange{Graphical Elements}} &
        \rotatebox[origin=l]{90}{\orange{Graphical Properties}} & \rotatebox[origin=l]{90}{\red{Spatial Navigation}} &
        \rotatebox[origin=l]{90}{\red{Hierarchy Drilling}} &
        \rotatebox[origin=l]{90}{\red{Multi-view}} &
        \rotatebox[origin=l]{90}{\green{Annotation}} &
        \rotatebox[origin=l]{90}{\green{Summarization}} &
        \rotatebox[origin=l]{90}{\green{Storytelling}} &
        \rotatebox[origin=l]{90}{Multi-turn Dialogue} &
        \rotatebox[origin=l]{90}{Multi-modality} \\
\hline
\multirowcell{65}{\rot{Systems and Tools}} & RGVisNet~\cite{song2022RGVisNetHybridRetrievalGeneration} \hfill 2022 & \levelone &  & \sand{\boldcheck} &  &  &  &  & \blue{\boldcheck} & \orange{\boldcheck} & \orange{\boldcheck} &  &  &  &  &  &  &  &  & \\
 & \cellcolor{gray!10}Sevi~\cite{tang2022SeviSpeechtoVisualizationNeural} \hfill 2022 & \cellcolor{gray!10}\levelone & \cellcolor{gray!10} & \cellcolor{gray!10}\sand{\boldcheck} & \cellcolor{gray!10} & \cellcolor{gray!10} & \cellcolor{gray!10} & \cellcolor{gray!10} & \cellcolor{gray!10}\blue{\boldcheck} & \cellcolor{gray!10}\orange{\boldcheck} & \cellcolor{gray!10}\orange{\boldcheck} & \cellcolor{gray!10} & \cellcolor{gray!10} & \cellcolor{gray!10} & \cellcolor{gray!10} & \cellcolor{gray!10} & \cellcolor{gray!10} & \cellcolor{gray!10} & \cellcolor{gray!10} & \cellcolor{gray!10}\boldcheck \\
 & ncNet~\cite{luo2022NaturalLanguageVisualization} \hfill 2022 & \levelone &  & \sand{\boldcheck} &  &  &  &  & \blue{\boldcheck} & \orange{\boldcheck} & \orange{\boldcheck} &  &  &  &  &  &  &  &  & \\
 & \cellcolor{gray!10}NL4DV 2~\cite{mitra2022FacilitatingConversationalInteraction} \hfill 2022 & \cellcolor{gray!10}\leveltwo & \cellcolor{gray!10}\purple{\boldcheck} & \cellcolor{gray!10}\sand{\boldcheck} & \cellcolor{gray!10} & \cellcolor{gray!10}\dawn{\boldcheck} & \cellcolor{gray!10} & \cellcolor{gray!10} & \cellcolor{gray!10}\blue{\boldcheck} & \cellcolor{gray!10}\orange{\boldcheck} & \cellcolor{gray!10}\orange{\boldcheck} & \cellcolor{gray!10} & \cellcolor{gray!10} & \cellcolor{gray!10}\red{\boldcheck} & \cellcolor{gray!10} & \cellcolor{gray!10} & \cellcolor{gray!10} & \cellcolor{gray!10} & \cellcolor{gray!10}\boldcheck & \cellcolor{gray!10}\boldcheck \\
 & PI2~\cite{chen2022PI2EndtoendInteractive} \hfill 2022 & \leveltwo &  & \sand{\boldcheck} &  & \dawn{\boldcheck} &  & \blue{\boldcheck} & \blue{\boldcheck} & \orange{\boldcheck} & \orange{\boldcheck} & \orange{\boldcheck} & \red{\boldcheck} & \red{\boldcheck} & \red{\boldcheck} & \green{\boldcheck} &  &  &  & \boldcheck \\
 & \cellcolor{gray!10}VIST5~\cite{voigt2023VIST5AdaptiveRetrievalAugmented} \hfill 2023 & \cellcolor{gray!10}\levelone & \cellcolor{gray!10} & \cellcolor{gray!10}\sand{\boldcheck} & \cellcolor{gray!10} & \cellcolor{gray!10}\dawn{\boldcheck} & \cellcolor{gray!10} & \cellcolor{gray!10} & \cellcolor{gray!10}\blue{\boldcheck} & \cellcolor{gray!10}\orange{\boldcheck} & \cellcolor{gray!10}\orange{\boldcheck} & \cellcolor{gray!10} & \cellcolor{gray!10}\red{\boldcheck} & \cellcolor{gray!10} & \cellcolor{gray!10}\red{\boldcheck} & \cellcolor{gray!10}\green{\boldcheck} & \cellcolor{gray!10} & \cellcolor{gray!10} & \cellcolor{gray!10}\boldcheck & \cellcolor{gray!10}\boldcheck \\
 & AI Threads~\cite{hong2023ConversationalAIThreads} \hfill 2023 & \levelone &  & \sand{\boldcheck} &  & \dawn{\boldcheck} & \blue{\boldcheck} &  & \blue{\boldcheck} & \orange{\boldcheck} & \orange{\boldcheck} & \orange{\boldcheck} &  &  &  & \green{\boldcheck} & \green{\boldcheck} &  & \boldcheck & \boldcheck \\
 & \cellcolor{gray!10}Chat2VIS~\cite{maddigan2023Chat2VISGeneratingData} \hfill 2023 & \cellcolor{gray!10}\levelone & \cellcolor{gray!10} & \cellcolor{gray!10}\sand{\boldcheck} & \cellcolor{gray!10} & \cellcolor{gray!10} & \cellcolor{gray!10}\blue{\boldcheck} & \cellcolor{gray!10} & \cellcolor{gray!10}\blue{\boldcheck} & \cellcolor{gray!10}\orange{\boldcheck} & \cellcolor{gray!10}\orange{\boldcheck} & \cellcolor{gray!10}\orange{\boldcheck} & \cellcolor{gray!10} & \cellcolor{gray!10} & \cellcolor{gray!10} & \cellcolor{gray!10} & \cellcolor{gray!10} & \cellcolor{gray!10} & \cellcolor{gray!10} & \cellcolor{gray!10} \\
 & DataTales~\cite{sultanum2023DATATALESInvestigatingUse} \hfill 2023 & \leveltwo & \purple{\boldcheck} & \sand{\boldcheck} &  & \dawn{\boldcheck} &  &  &  &  &  &  &  &  &  & \green{\boldcheck} & \green{\boldcheck} & \green{\boldcheck} & \boldcheck & \boldcheck \\
 & \cellcolor{gray!10}LIDA~\cite{dibia2023LIDAToolAutomatic} \hfill 2023 & \cellcolor{gray!10}\leveltwo & \cellcolor{gray!10}\purple{\boldcheck} & \cellcolor{gray!10}\sand{\boldcheck} & \cellcolor{gray!10} & \cellcolor{gray!10}\dawn{\boldcheck} & \cellcolor{gray!10}\blue{\boldcheck} & \cellcolor{gray!10} & \cellcolor{gray!10}\blue{\boldcheck} & \cellcolor{gray!10}\orange{\boldcheck} & \cellcolor{gray!10}\orange{\boldcheck} & \cellcolor{gray!10}\orange{\boldcheck} & \cellcolor{gray!10} & \cellcolor{gray!10} & \cellcolor{gray!10} & \cellcolor{gray!10} & \cellcolor{gray!10} & \cellcolor{gray!10} & \cellcolor{gray!10}\boldcheck & \cellcolor{gray!10}\boldcheck \\
 & Text2Chart31~\cite{pesaranzadeh2024Text2Chart31InstructionTuning} \hfill 2024 & \levelone &  & \sand{\boldcheck} &  &  &  &  & \blue{\boldcheck} & \orange{\boldcheck} & \orange{\boldcheck} & \orange{\boldcheck} &  &  &  &  &  &  &  & \\
 & \cellcolor{gray!10}Wu et al.~\cite{wu2024AutomatedDataVisualization} \hfill 2024 & \cellcolor{gray!10}\levelone & \cellcolor{gray!10} & \cellcolor{gray!10}\sand{\boldcheck} & \cellcolor{gray!10} & \cellcolor{gray!10} & \cellcolor{gray!10} & \cellcolor{gray!10}\blue{\boldcheck} & \cellcolor{gray!10}\blue{\boldcheck} & \cellcolor{gray!10}\orange{\boldcheck} & \cellcolor{gray!10}\orange{\boldcheck} & \cellcolor{gray!10}\orange{\boldcheck} & \cellcolor{gray!10} & \cellcolor{gray!10} & \cellcolor{gray!10} & \cellcolor{gray!10} & \cellcolor{gray!10} & \cellcolor{gray!10} & \cellcolor{gray!10} & \cellcolor{gray!10} \\
 & Li et al.~\cite{li2024VisualizationGenerationLarge} \hfill 2024 & \levelone &  & \sand{\boldcheck} &  &  &  & \blue{\boldcheck} & \blue{\boldcheck} & \orange{\boldcheck} & \orange{\boldcheck} & \orange{\boldcheck} &  &  &  &  &  &  &  & \\
 & \cellcolor{gray!10}Liu et al.~\cite{liu2024BreathingNewLife} \hfill 2024 & \cellcolor{gray!10}\levelone & \cellcolor{gray!10} & \cellcolor{gray!10}\sand{\boldcheck} & \cellcolor{gray!10} & \cellcolor{gray!10}\dawn{\boldcheck} & \cellcolor{gray!10} & \cellcolor{gray!10} & \cellcolor{gray!10}\blue{\boldcheck} & \cellcolor{gray!10}\orange{\boldcheck} & \cellcolor{gray!10}\orange{\boldcheck} & \cellcolor{gray!10}\orange{\boldcheck} & \cellcolor{gray!10} & \cellcolor{gray!10} & \cellcolor{gray!10} & \cellcolor{gray!10}\green{\boldcheck} & \cellcolor{gray!10} & \cellcolor{gray!10} & \cellcolor{gray!10}\boldcheck & \cellcolor{gray!10} \\
 & PhenoFlow~\cite{kim2024PhenoFlowHumanLLMDriven} \hfill 2024 & \leveltwo & \purple{\boldcheck} & \sand{\boldcheck} &  & \dawn{\boldcheck} & \blue{\boldcheck} & \blue{\boldcheck} & \blue{\boldcheck} & \orange{\boldcheck} & \orange{\boldcheck} &  & \red{\boldcheck} & \red{\boldcheck} & \red{\boldcheck} &  &  &  & \boldcheck & \boldcheck \\
 & \cellcolor{gray!10}Blace~\cite{lyi2024LearnableExpressiveVisualization} \hfill 2024 & \cellcolor{gray!10}\leveltwo & \cellcolor{gray!10} & \cellcolor{gray!10}\sand{\boldcheck} & \cellcolor{gray!10} & \cellcolor{gray!10}\dawn{\boldcheck} & \cellcolor{gray!10} & \cellcolor{gray!10} & \cellcolor{gray!10}\blue{\boldcheck} & \cellcolor{gray!10}\orange{\boldcheck} & \cellcolor{gray!10}\orange{\boldcheck} & \cellcolor{gray!10} & \cellcolor{gray!10}\red{\boldcheck} & \cellcolor{gray!10} & \cellcolor{gray!10} & \cellcolor{gray!10} & \cellcolor{gray!10} & \cellcolor{gray!10} & \cellcolor{gray!10} & \cellcolor{gray!10} \\
 & LEVA~\cite{zhao2024LEVAUsingLarge} \hfill 2024 & \leveltwo & \purple{\boldcheck} & \sand{\boldcheck} &  & \dawn{\boldcheck} &  &  & \blue{\boldcheck} & \orange{\boldcheck} & \orange{\boldcheck} & \orange{\boldcheck} & \red{\boldcheck} & \red{\boldcheck} & \red{\boldcheck} & \green{\boldcheck} & \green{\boldcheck} & \green{\boldcheck} & \boldcheck & \boldcheck \\
 & \cellcolor{gray!10}Bavisitter~\cite{choi2024BavisitterIntegratingDesign} \hfill 2024 & \cellcolor{gray!10}\leveltwo & \cellcolor{gray!10} & \cellcolor{gray!10}\sand{\boldcheck} & \cellcolor{gray!10}\cyan{\boldcheck} & \cellcolor{gray!10}\dawn{\boldcheck} & \cellcolor{gray!10} & \cellcolor{gray!10} & \cellcolor{gray!10}\blue{\boldcheck} & \cellcolor{gray!10}\orange{\boldcheck} & \cellcolor{gray!10}\orange{\boldcheck} & \cellcolor{gray!10}\orange{\boldcheck} & \cellcolor{gray!10} & \cellcolor{gray!10} & \cellcolor{gray!10} & \cellcolor{gray!10}\green{\boldcheck} & \cellcolor{gray!10} & \cellcolor{gray!10} & \cellcolor{gray!10}\boldcheck & \cellcolor{gray!10}\boldcheck \\
 & MMCoVisNet~\cite{song2024MarryingDialogueSystems} \hfill 2024 & \leveltwo & \purple{\boldcheck} & \sand{\boldcheck} &  & \dawn{\boldcheck} & \blue{\boldcheck} &  & \blue{\boldcheck} & \orange{\boldcheck} & \orange{\boldcheck} &  &  &  &  &  &  &  & \boldcheck & \\
 & \cellcolor{gray!10}DynaVis~\cite{vaithilingam2024DynaVisDynamicallySynthesized} \hfill 2024 & \cellcolor{gray!10}\leveltwo & \cellcolor{gray!10}\purple{\boldcheck} & \cellcolor{gray!10}\sand{\boldcheck} & \cellcolor{gray!10} & \cellcolor{gray!10}\dawn{\boldcheck} & \cellcolor{gray!10}\blue{\boldcheck} & \cellcolor{gray!10} & \cellcolor{gray!10}\blue{\boldcheck} & \cellcolor{gray!10}\orange{\boldcheck} & \cellcolor{gray!10}\orange{\boldcheck} & \cellcolor{gray!10}\orange{\boldcheck} & \cellcolor{gray!10} & \cellcolor{gray!10} & \cellcolor{gray!10} & \cellcolor{gray!10}\green{\boldcheck} & \cellcolor{gray!10}\green{\boldcheck} & \cellcolor{gray!10} & \cellcolor{gray!10}\boldcheck & \cellcolor{gray!10}\boldcheck \\
 & Data Formulator~\cite{wang2023DataFormulatorAIpowered} \hfill 2024 & \leveltwo &  & \sand{\boldcheck} &  &  &  &  & \blue{\boldcheck} & \orange{\boldcheck} & \orange{\boldcheck} & \orange{\boldcheck} &  &  &  &  &  &  &  & \boldcheck \\
 & \cellcolor{gray!10}DASH~\cite{bromley2024DASHBimodalData} \hfill 2024 & \cellcolor{gray!10}\leveltwo & \cellcolor{gray!10}\purple{\boldcheck} & \cellcolor{gray!10}\sand{\boldcheck} & \cellcolor{gray!10} & \cellcolor{gray!10}\dawn{\boldcheck} & \cellcolor{gray!10}\blue{\boldcheck} & \cellcolor{gray!10} & \cellcolor{gray!10}\blue{\boldcheck} & \cellcolor{gray!10}\orange{\boldcheck} & \cellcolor{gray!10}\orange{\boldcheck} & \cellcolor{gray!10}\orange{\boldcheck} & \cellcolor{gray!10} & \cellcolor{gray!10}\red{\boldcheck} & \cellcolor{gray!10}\red{\boldcheck} & \cellcolor{gray!10}\green{\boldcheck} & \cellcolor{gray!10}\green{\boldcheck} & \cellcolor{gray!10}\green{\boldcheck} & \cellcolor{gray!10} & \cellcolor{gray!10}\boldcheck \\
 & NL4DV-LLM~\cite{sah2024GeneratingAnalyticSpecifications} \hfill 2024 & \levelthree & \purple{\boldcheck} & \sand{\boldcheck} & \cyan{\boldcheck} & \dawn{\boldcheck} &  &  & \blue{\boldcheck} & \orange{\boldcheck} & \orange{\boldcheck} & \orange{\boldcheck} &  &  &  &  &  &  & \boldcheck & \\
 & \cellcolor{gray!10}FathomGPT~\cite{khanal2024FathomGPTNaturalLanguage} \hfill 2024 & \cellcolor{gray!10}\levelthree & \cellcolor{gray!10}\purple{\boldcheck} & \cellcolor{gray!10}\sand{\boldcheck} & \cellcolor{gray!10}\cyan{\boldcheck} & \cellcolor{gray!10}\dawn{\boldcheck} & \cellcolor{gray!10}\blue{\boldcheck} & \cellcolor{gray!10}\blue{\boldcheck} & \cellcolor{gray!10}\blue{\boldcheck} & \cellcolor{gray!10}\orange{\boldcheck} & \cellcolor{gray!10}\orange{\boldcheck} & \cellcolor{gray!10}\orange{\boldcheck} & \cellcolor{gray!10}\red{\boldcheck} & \cellcolor{gray!10}\red{\boldcheck} & \cellcolor{gray!10}\red{\boldcheck} & \cellcolor{gray!10}\green{\boldcheck} & \cellcolor{gray!10}\green{\boldcheck} & \cellcolor{gray!10} & \cellcolor{gray!10}\boldcheck & \cellcolor{gray!10}\boldcheck \\
 & MatPlotAgent~\cite{yang2024MatPlotAgentMethodEvaluation} \hfill 2024 & \levelthree & \purple{\boldcheck} & \sand{\boldcheck} & \cyan{\boldcheck} & \dawn{\boldcheck} &  &  & \blue{\boldcheck} & \orange{\boldcheck} & \orange{\boldcheck} & \orange{\boldcheck} &  &  &  & \green{\boldcheck} &  &  &  & \boldcheck \\
 & \cellcolor{gray!10}AVA~\cite{ava2024} \hfill 2024 & \cellcolor{gray!10}\levelthree & \cellcolor{gray!10}\purple{\boldcheck} & \cellcolor{gray!10}\sand{\boldcheck} & \cellcolor{gray!10}\cyan{\boldcheck} & \cellcolor{gray!10}\dawn{\boldcheck} & \cellcolor{gray!10} & \cellcolor{gray!10} & \cellcolor{gray!10}\blue{\boldcheck} & \cellcolor{gray!10}\orange{\boldcheck} & \cellcolor{gray!10}\orange{\boldcheck} & \cellcolor{gray!10}\orange{\boldcheck} & \cellcolor{gray!10}\red{\boldcheck} & \cellcolor{gray!10} & \cellcolor{gray!10} & \cellcolor{gray!10} & \cellcolor{gray!10} & \cellcolor{gray!10} & \cellcolor{gray!10}\boldcheck & \cellcolor{gray!10}\boldcheck \\
 & DataVisT5~\cite{wan2025DataVisT5PretrainedLanguage} \hfill 2025 & \levelone &  & \sand{\boldcheck} &  &  &  & \blue{\boldcheck} & \blue{\boldcheck} & \orange{\boldcheck} & \orange{\boldcheck} & \orange{\boldcheck} &  &  &  &  & \green{\boldcheck} &  &  & \boldcheck \\
 & \cellcolor{gray!10}HARVis~\cite{shi2025AugmentingRealisticCharts} \hfill 2025 & \cellcolor{gray!10}\levelone & \cellcolor{gray!10} & \cellcolor{gray!10}\sand{\boldcheck} & \cellcolor{gray!10} & \cellcolor{gray!10}\dawn{\boldcheck} & \cellcolor{gray!10} & \cellcolor{gray!10} & \cellcolor{gray!10}\blue{\boldcheck} & \cellcolor{gray!10}\orange{\boldcheck} & \cellcolor{gray!10}\orange{\boldcheck} & \cellcolor{gray!10}\orange{\boldcheck} & \cellcolor{gray!10}\red{\boldcheck} & \cellcolor{gray!10} & \cellcolor{gray!10} & \cellcolor{gray!10}\green{\boldcheck} & \cellcolor{gray!10} & \cellcolor{gray!10} & \cellcolor{gray!10} & \cellcolor{gray!10}\boldcheck \\
 & SpeechVisNet~\cite{zhang2026SpeechtoVisualizationEndtoEndSpeechDriven} \hfill 2025 & \levelone &  & \sand{\boldcheck} &  &  &  &  & \blue{\boldcheck} & \orange{\boldcheck} & \orange{\boldcheck} &  &  &  &  &  &  &  &  & \boldcheck \\
 & \cellcolor{gray!10}CoML4Vis~\cite{chen2025VisEvalBenchmarkData} \hfill 2025 & \cellcolor{gray!10}\levelone & \cellcolor{gray!10} & \cellcolor{gray!10}\sand{\boldcheck} & \cellcolor{gray!10} & \cellcolor{gray!10} & \cellcolor{gray!10}\blue{\boldcheck} & \cellcolor{gray!10}\blue{\boldcheck} & \cellcolor{gray!10}\blue{\boldcheck} & \cellcolor{gray!10}\orange{\boldcheck} & \cellcolor{gray!10}\orange{\boldcheck} & \cellcolor{gray!10}\orange{\boldcheck} & \cellcolor{gray!10} & \cellcolor{gray!10} & \cellcolor{gray!10} & \cellcolor{gray!10} & \cellcolor{gray!10} & \cellcolor{gray!10} & \cellcolor{gray!10} & \cellcolor{gray!10} \\
 & Shao et al.~\cite{shao2025dolanguagemodelagents} \hfill 2025 & \levelone &  &  & \cyan{\boldcheck} &  &  &  &  & \orange{\boldcheck} & \orange{\boldcheck} & \orange{\boldcheck} &  &  &  &  & \green{\boldcheck} &  &  & \\
 & \cellcolor{gray!10}NL4DV-Stylist~\cite{ji2025NL4DVStylistStylingData} \hfill 2025 & \cellcolor{gray!10}\leveltwo & \cellcolor{gray!10}\purple{\boldcheck} & \cellcolor{gray!10}\sand{\boldcheck} & \cellcolor{gray!10} & \cellcolor{gray!10}\dawn{\boldcheck} & \cellcolor{gray!10} & \cellcolor{gray!10} & \cellcolor{gray!10}\blue{\boldcheck} & \cellcolor{gray!10}\orange{\boldcheck} & \cellcolor{gray!10}\orange{\boldcheck} & \cellcolor{gray!10}\orange{\boldcheck} & \cellcolor{gray!10} & \cellcolor{gray!10} & \cellcolor{gray!10} & \cellcolor{gray!10} & \cellcolor{gray!10} & \cellcolor{gray!10} & \cellcolor{gray!10}\boldcheck & \cellcolor{gray!10}\boldcheck \\
 & Prompt4Vis~\cite{li2024Prompt4VisPromptingLarge} \hfill 2025 & \leveltwo & \purple{\boldcheck} & \sand{\boldcheck} &  & \dawn{\boldcheck} & \blue{\boldcheck} & \blue{\boldcheck} & \blue{\boldcheck} & \orange{\boldcheck} & \orange{\boldcheck} & \orange{\boldcheck} &  &  &  &  &  &  &  & \\
 & \cellcolor{gray!10}SUPQA~\cite{huang2025SUPQALLMbasedGeoVisualization} \hfill 2025 & \cellcolor{gray!10}\leveltwo & \cellcolor{gray!10} & \cellcolor{gray!10}\sand{\boldcheck} & \cellcolor{gray!10} & \cellcolor{gray!10}\dawn{\boldcheck} & \cellcolor{gray!10}\blue{\boldcheck} & \cellcolor{gray!10}\blue{\boldcheck} & \cellcolor{gray!10}\blue{\boldcheck} & \cellcolor{gray!10}\orange{\boldcheck} & \cellcolor{gray!10}\orange{\boldcheck} & \cellcolor{gray!10}\orange{\boldcheck} & \cellcolor{gray!10}\red{\boldcheck} & \cellcolor{gray!10}\red{\boldcheck} & \cellcolor{gray!10}\red{\boldcheck} & \cellcolor{gray!10}\green{\boldcheck} & \cellcolor{gray!10}\green{\boldcheck} & \cellcolor{gray!10} & \cellcolor{gray!10} & \cellcolor{gray!10}\boldcheck \\
 & Aggarwal et al.~\cite{aggarwal2025GoalDrivenDataStory} \hfill 2025 & \leveltwo & \purple{\boldcheck} & \sand{\boldcheck} &  & \dawn{\boldcheck} &  &  & \blue{\boldcheck} & \orange{\boldcheck} & \orange{\boldcheck} & \orange{\boldcheck} &  & \red{\boldcheck} & \red{\boldcheck} &  & \green{\boldcheck} & \green{\boldcheck} &  & \boldcheck \\
 & \cellcolor{gray!10}Data Formulator 2~\cite{wang2025DataFormulator2} \hfill 2025 & \cellcolor{gray!10}\leveltwo & \cellcolor{gray!10}\purple{\boldcheck} & \cellcolor{gray!10}\sand{\boldcheck} & \cellcolor{gray!10} & \cellcolor{gray!10}\dawn{\boldcheck} & \cellcolor{gray!10} & \cellcolor{gray!10} & \cellcolor{gray!10}\blue{\boldcheck} & \cellcolor{gray!10}\orange{\boldcheck} & \cellcolor{gray!10}\orange{\boldcheck} & \cellcolor{gray!10}\orange{\boldcheck} & \cellcolor{gray!10} & \cellcolor{gray!10}\red{\boldcheck} & \cellcolor{gray!10} & \cellcolor{gray!10}\green{\boldcheck} & \cellcolor{gray!10} & \cellcolor{gray!10} & \cellcolor{gray!10}\boldcheck & \cellcolor{gray!10}\boldcheck \\
 & Pluto~\cite{srinivasan2025PlutoAuthoringSemantically} \hfill 2025 & \leveltwo &  & \sand{\boldcheck} &  & \dawn{\boldcheck} &  &  & \blue{\boldcheck} & \orange{\boldcheck} & \orange{\boldcheck} & \orange{\boldcheck} &  &  &  & \green{\boldcheck} & \green{\boldcheck} & \green{\boldcheck} & \boldcheck & \boldcheck \\
 & \cellcolor{gray!10}RECITKIT~\cite{setlur2025RECITKITSpatialToolkit} \hfill 2025 & \cellcolor{gray!10}\leveltwo & \cellcolor{gray!10}\purple{\boldcheck} & \cellcolor{gray!10}\sand{\boldcheck} & \cellcolor{gray!10} & \cellcolor{gray!10}\dawn{\boldcheck} & \cellcolor{gray!10} & \cellcolor{gray!10}\blue{\boldcheck} & \cellcolor{gray!10}\blue{\boldcheck} & \cellcolor{gray!10}\orange{\boldcheck} & \cellcolor{gray!10}\orange{\boldcheck} & \cellcolor{gray!10}\orange{\boldcheck} & \cellcolor{gray!10}\red{\boldcheck} & \cellcolor{gray!10}\red{\boldcheck} & \cellcolor{gray!10}\red{\boldcheck} & \cellcolor{gray!10}\green{\boldcheck} & \cellcolor{gray!10} & \cellcolor{gray!10}\green{\boldcheck} & \cellcolor{gray!10} & \cellcolor{gray!10}\boldcheck \\
 & Step-NL2VIS~\cite{luo2025NvBench20Resolving} \hfill 2025 & \leveltwo &  & \sand{\boldcheck} &  & \dawn{\boldcheck} &  &  & \blue{\boldcheck} & \orange{\boldcheck} & \orange{\boldcheck} & \orange{\boldcheck} &  &  &  &  &  &  &  & \\
 & \cellcolor{gray!10}InterChat~\cite{chen2025InterChatEnhancingGenerative} \hfill 2025 & \cellcolor{gray!10}\leveltwo & \cellcolor{gray!10}\purple{\boldcheck} & \cellcolor{gray!10}\sand{\boldcheck} & \cellcolor{gray!10} & \cellcolor{gray!10}\dawn{\boldcheck} & \cellcolor{gray!10}\blue{\boldcheck} & \cellcolor{gray!10} & \cellcolor{gray!10}\blue{\boldcheck} & \cellcolor{gray!10}\orange{\boldcheck} & \cellcolor{gray!10}\orange{\boldcheck} & \cellcolor{gray!10}\orange{\boldcheck} & \cellcolor{gray!10}\red{\boldcheck} & \cellcolor{gray!10} & \cellcolor{gray!10}\red{\boldcheck} & \cellcolor{gray!10}\green{\boldcheck} & \cellcolor{gray!10}\green{\boldcheck} & \cellcolor{gray!10} & \cellcolor{gray!10}\boldcheck & \cellcolor{gray!10}\boldcheck \\
 & Dataweaver~\cite{fu2025DATAWEAVERAuthoringDataDriven} \hfill 2025 & \leveltwo & \purple{\boldcheck} & \sand{\boldcheck} &  & \dawn{\boldcheck} &  &  & \blue{\boldcheck} & \orange{\boldcheck} & \orange{\boldcheck} & \orange{\boldcheck} & \red{\boldcheck} & \red{\boldcheck} & \red{\boldcheck} & \green{\boldcheck} & \green{\boldcheck} & \green{\boldcheck} &  & \\
 & \cellcolor{gray!10}ChartGPT~\cite{tian2025ChartGPTLeveragingLLMs} \hfill 2025 & \cellcolor{gray!10}\leveltwo & \cellcolor{gray!10} & \cellcolor{gray!10}\sand{\boldcheck} & \cellcolor{gray!10} & \cellcolor{gray!10} & \cellcolor{gray!10} & \cellcolor{gray!10} & \cellcolor{gray!10}\blue{\boldcheck} & \cellcolor{gray!10}\orange{\boldcheck} & \cellcolor{gray!10}\orange{\boldcheck} & \cellcolor{gray!10} & \cellcolor{gray!10} & \cellcolor{gray!10} & \cellcolor{gray!10} & \cellcolor{gray!10} & \cellcolor{gray!10} & \cellcolor{gray!10} & \cellcolor{gray!10} & \cellcolor{gray!10}\boldcheck \\
 & VizTA~\cite{vizTA2025} \hfill 2025 & \leveltwo & \purple{\boldcheck} & \sand{\boldcheck} &  & \dawn{\boldcheck} &  &  & \blue{\boldcheck} &  &  & \orange{\boldcheck} &  &  &  & \green{\boldcheck} & \green{\boldcheck} &  & \boldcheck & \boldcheck \\
 & \cellcolor{gray!10}nvAgent~\cite{ouyang2025NvAgentAutomatedData} \hfill 2025 & \cellcolor{gray!10}\levelthree & \cellcolor{gray!10}\purple{\boldcheck} & \cellcolor{gray!10}\sand{\boldcheck} & \cellcolor{gray!10}\cyan{\boldcheck} & \cellcolor{gray!10}\dawn{\boldcheck} & \cellcolor{gray!10}\blue{\boldcheck} & \cellcolor{gray!10}\blue{\boldcheck} & \cellcolor{gray!10}\blue{\boldcheck} & \cellcolor{gray!10}\orange{\boldcheck} & \cellcolor{gray!10}\orange{\boldcheck} & \cellcolor{gray!10} & \cellcolor{gray!10} & \cellcolor{gray!10} & \cellcolor{gray!10} & \cellcolor{gray!10} & \cellcolor{gray!10} & \cellcolor{gray!10} & \cellcolor{gray!10} & \cellcolor{gray!10} \\
 & $C^2$~\cite{koh2025C^2ScalableAutoFeedback} \hfill 2025 & \levelthree & \purple{\boldcheck} & \sand{\boldcheck} & \cyan{\boldcheck} & \dawn{\boldcheck} &  &  & \blue{\boldcheck} & \orange{\boldcheck} & \orange{\boldcheck} & \orange{\boldcheck} &  &  &  & \green{\boldcheck} &  &  & \boldcheck & \\
 & \cellcolor{gray!10}Plume~\cite{lisnic2025PlumeScaffoldingText} \hfill 2025 & \cellcolor{gray!10}\levelthree & \cellcolor{gray!10}\purple{\boldcheck} & \cellcolor{gray!10}\sand{\boldcheck} & \cellcolor{gray!10}\cyan{\boldcheck} & \cellcolor{gray!10}\dawn{\boldcheck} & \cellcolor{gray!10} & \cellcolor{gray!10} & \cellcolor{gray!10} & \cellcolor{gray!10}\orange{\boldcheck} & \cellcolor{gray!10}\orange{\boldcheck} & \cellcolor{gray!10}\orange{\boldcheck} & \cellcolor{gray!10} & \cellcolor{gray!10} & \cellcolor{gray!10}\red{\boldcheck} & \cellcolor{gray!10}\green{\boldcheck} & \cellcolor{gray!10}\green{\boldcheck} & \cellcolor{gray!10}\green{\boldcheck} & \cellcolor{gray!10}\boldcheck & \cellcolor{gray!10} \\
 & Jupybara~\cite{wang2025JupybaraOperationalizingDesign} \hfill 2025 & \levelthree & \purple{\boldcheck} & \sand{\boldcheck} & \cyan{\boldcheck} & \dawn{\boldcheck} &  &  & \blue{\boldcheck} & \orange{\boldcheck} & \orange{\boldcheck} & \orange{\boldcheck} &  &  &  & \green{\boldcheck} & \green{\boldcheck} & \green{\boldcheck} & \boldcheck & \boldcheck \\
 & \cellcolor{gray!10}DeepVis~\cite{shuai2025DeepVISBridgingNatural} \hfill 2025 & \cellcolor{gray!10}\levelthree & \cellcolor{gray!10}\purple{\boldcheck} & \cellcolor{gray!10}\sand{\boldcheck} & \cellcolor{gray!10}\cyan{\boldcheck} & \cellcolor{gray!10}\dawn{\boldcheck} & \cellcolor{gray!10}\blue{\boldcheck} & \cellcolor{gray!10} & \cellcolor{gray!10}\blue{\boldcheck} & \cellcolor{gray!10}\orange{\boldcheck} & \cellcolor{gray!10}\orange{\boldcheck} & \cellcolor{gray!10} & \cellcolor{gray!10} & \cellcolor{gray!10} & \cellcolor{gray!10} & \cellcolor{gray!10} & \cellcolor{gray!10}\green{\boldcheck} & \cellcolor{gray!10} & \cellcolor{gray!10} & \cellcolor{gray!10}\boldcheck \\
 & Text2Vis~\cite{rahman2025Text2VisChallengingDiverse} \hfill 2025 & \levelthree &  & \sand{\boldcheck} & \cyan{\boldcheck} &  &  &  & \blue{\boldcheck} & \orange{\boldcheck} & \orange{\boldcheck} & \orange{\boldcheck} &  &  & \red{\boldcheck} & \green{\boldcheck} &  &  &  & \\
 & \cellcolor{gray!10}DataNarrative~\cite{islam2024DataNarrativeAutomatedDataDriven} \hfill 2025 & \cellcolor{gray!10}\levelthree & \cellcolor{gray!10}\purple{\boldcheck} & \cellcolor{gray!10}\sand{\boldcheck} & \cellcolor{gray!10}\cyan{\boldcheck} & \cellcolor{gray!10}\dawn{\boldcheck} & \cellcolor{gray!10}\blue{\boldcheck} & \cellcolor{gray!10} & \cellcolor{gray!10} & \cellcolor{gray!10}\orange{\boldcheck} & \cellcolor{gray!10}\orange{\boldcheck} & \cellcolor{gray!10}\orange{\boldcheck} & \cellcolor{gray!10} & \cellcolor{gray!10} & \cellcolor{gray!10} & \cellcolor{gray!10}\green{\boldcheck} & \cellcolor{gray!10} & \cellcolor{gray!10}\green{\boldcheck} & \cellcolor{gray!10} & \cellcolor{gray!10} \\
 & CoDA~\cite{chen2025codaagenticsystemscollaborative} \hfill 2025 & \levelthree & \purple{\boldcheck} & \sand{\boldcheck} & \cyan{\boldcheck} & \dawn{\boldcheck} & \blue{\boldcheck} &  & \blue{\boldcheck} & \orange{\boldcheck} & \orange{\boldcheck} & \orange{\boldcheck} &  &  & \red{\boldcheck} &  &  &  &  & \\
 & \cellcolor{gray!10}Gyarmati et al.~\cite{gyarmati2025composableagenticautomatedvisual} \hfill 2025 & \cellcolor{gray!10}\levelthree & \cellcolor{gray!10}\purple{\boldcheck} & \cellcolor{gray!10}\sand{\boldcheck} & \cellcolor{gray!10}\cyan{\boldcheck} & \cellcolor{gray!10}\dawn{\boldcheck} & \cellcolor{gray!10}\blue{\boldcheck} & \cellcolor{gray!10} & \cellcolor{gray!10}\blue{\boldcheck} & \cellcolor{gray!10}\orange{\boldcheck} & \cellcolor{gray!10}\orange{\boldcheck} & \cellcolor{gray!10}\orange{\boldcheck} & \cellcolor{gray!10} & \cellcolor{gray!10} & \cellcolor{gray!10} & \cellcolor{gray!10} & \cellcolor{gray!10}\green{\boldcheck} & \cellcolor{gray!10}\green{\boldcheck} & \cellcolor{gray!10} & \cellcolor{gray!10} \\
 & LightVA~\cite{zhao2025LightVALightweightVisual} \hfill 2025 & \levelfour & \purple{\boldcheck} & \sand{\boldcheck} & \cyan{\boldcheck} & \dawn{\boldcheck} & \blue{\boldcheck} &  & \blue{\boldcheck} & \orange{\boldcheck} & \orange{\boldcheck} & \orange{\boldcheck} & \red{\boldcheck} & \red{\boldcheck} & \red{\boldcheck} & \green{\boldcheck} & \green{\boldcheck} &  & \boldcheck & \boldcheck \\
 & \cellcolor{gray!10}GeoAuthor~\cite{chen2026GeoauthorLinkingTextVisualization} \hfill 2026 & \cellcolor{gray!10}\levelone & \cellcolor{gray!10} & \cellcolor{gray!10}\sand{\boldcheck} & \cellcolor{gray!10} & \cellcolor{gray!10}\dawn{\boldcheck} & \cellcolor{gray!10} & \cellcolor{gray!10} & \cellcolor{gray!10}\blue{\boldcheck} & \cellcolor{gray!10}\orange{\boldcheck} & \cellcolor{gray!10}\orange{\boldcheck} & \cellcolor{gray!10}\orange{\boldcheck} & \cellcolor{gray!10}\red{\boldcheck} & \cellcolor{gray!10} & \cellcolor{gray!10} & \cellcolor{gray!10}\green{\boldcheck} & \cellcolor{gray!10}\green{\boldcheck} & \cellcolor{gray!10}\green{\boldcheck} & \cellcolor{gray!10} & \cellcolor{gray!10}\boldcheck \\
 & Lu et al.~\cite{lu2026RobustInteractiveTextVisualization} \hfill 2026 & \levelone &  & \sand{\boldcheck} &  &  &  &  & \blue{\boldcheck} & \orange{\boldcheck} & \orange{\boldcheck} & \orange{\boldcheck} &  &  &  &  &  &  & \boldcheck & \\
 & \cellcolor{gray!10}Tytarenko et al.~\cite{tytarenko2026ConfirmationBiasAwarenessVisual} \hfill 2026 & \cellcolor{gray!10}\levelone & \cellcolor{gray!10} & \cellcolor{gray!10} & \cellcolor{gray!10}\cyan{\boldcheck} & \cellcolor{gray!10}\dawn{\boldcheck} & \cellcolor{gray!10} & \cellcolor{gray!10} & \cellcolor{gray!10} & \cellcolor{gray!10} & \cellcolor{gray!10} & \cellcolor{gray!10} & \cellcolor{gray!10} & \cellcolor{gray!10} & \cellcolor{gray!10} & \cellcolor{gray!10} & \cellcolor{gray!10} & \cellcolor{gray!10} & \cellcolor{gray!10}\boldcheck & \cellcolor{gray!10}\boldcheck \\
 & Int.\ Drill-Down~\cite{zheng2026IntelligentDrillDownLarge} \hfill 2026 & \leveltwo & \purple{\boldcheck} & \sand{\boldcheck} &  & \dawn{\boldcheck} & \blue{\boldcheck} &  & \blue{\boldcheck} & \orange{\boldcheck} & \orange{\boldcheck} & \orange{\boldcheck} & \red{\boldcheck} & \red{\boldcheck} &  &  & \green{\boldcheck} &  & \boldcheck & \boldcheck \\
 & \cellcolor{gray!10}Vis.\ Autocomplete~\cite{jeon2026VisualizationAutocompleteVisualizationAuthoring} \hfill 2026 & \cellcolor{gray!10}\leveltwo & \cellcolor{gray!10}\purple{\boldcheck} & \cellcolor{gray!10}\sand{\boldcheck} & \cellcolor{gray!10} & \cellcolor{gray!10}\dawn{\boldcheck} & \cellcolor{gray!10} & \cellcolor{gray!10} & \cellcolor{gray!10}\blue{\boldcheck} & \cellcolor{gray!10}\orange{\boldcheck} & \cellcolor{gray!10}\orange{\boldcheck} & \cellcolor{gray!10}\orange{\boldcheck} & \cellcolor{gray!10} & \cellcolor{gray!10} & \cellcolor{gray!10} & \cellcolor{gray!10} & \cellcolor{gray!10} & \cellcolor{gray!10} & \cellcolor{gray!10}\boldcheck & \cellcolor{gray!10} \\
 & Vistoryteller~\cite{shi2026VistorytellerDesigningDataStories} \hfill 2026 & \levelthree & \purple{\boldcheck} & \sand{\boldcheck} & \cyan{\boldcheck} & \dawn{\boldcheck} & \blue{\boldcheck} &  & \blue{\boldcheck} & \orange{\boldcheck} & \orange{\boldcheck} & \orange{\boldcheck} &  &  &  & \green{\boldcheck} & \green{\boldcheck} & \green{\boldcheck} & \boldcheck & \boldcheck \\
 & \cellcolor{gray!10}SASAV~\cite{sun2026SasavSelfDirectedAgent} \hfill 2026 & \cellcolor{gray!10}\levelthree & \cellcolor{gray!10}\purple{\boldcheck} & \cellcolor{gray!10}\sand{\boldcheck} & \cellcolor{gray!10}\cyan{\boldcheck} & \cellcolor{gray!10}\dawn{\boldcheck} & \cellcolor{gray!10}\blue{\boldcheck} & \cellcolor{gray!10} & \cellcolor{gray!10}\blue{\boldcheck} & \cellcolor{gray!10}\orange{\boldcheck} & \cellcolor{gray!10}\orange{\boldcheck} & \cellcolor{gray!10}\orange{\boldcheck} & \cellcolor{gray!10}\red{\boldcheck} & \cellcolor{gray!10} & \cellcolor{gray!10} & \cellcolor{gray!10} & \cellcolor{gray!10}\green{\boldcheck} & \cellcolor{gray!10} & \cellcolor{gray!10} & \cellcolor{gray!10}\boldcheck \\
 & BONSAI~\cite{spinner2026BonsaiMixedInitiativeWorkspace} \hfill 2026 & \levelthree & \purple{\boldcheck} & \sand{\boldcheck} & \cyan{\boldcheck} & \dawn{\boldcheck} &  &  & \blue{\boldcheck} & \orange{\boldcheck} & \orange{\boldcheck} & \orange{\boldcheck} &  &  & \red{\boldcheck} &  &  &  & \boldcheck & \boldcheck \\
 & \cellcolor{gray!10}FlowForge~\cite{flowforge2026} \hfill 2026 & \cellcolor{gray!10}\levelfour & \cellcolor{gray!10}\purple{\boldcheck} & \cellcolor{gray!10}\sand{\boldcheck} & \cellcolor{gray!10}\cyan{\boldcheck} & \cellcolor{gray!10}\dawn{\boldcheck} & \cellcolor{gray!10}\blue{\boldcheck} & \cellcolor{gray!10} & \cellcolor{gray!10}\blue{\boldcheck} & \cellcolor{gray!10}\orange{\boldcheck} & \cellcolor{gray!10}\orange{\boldcheck} & \cellcolor{gray!10}\orange{\boldcheck} & \cellcolor{gray!10}\red{\boldcheck} & \cellcolor{gray!10}\red{\boldcheck} & \cellcolor{gray!10}\red{\boldcheck} & \cellcolor{gray!10} & \cellcolor{gray!10}\green{\boldcheck} & \cellcolor{gray!10}\green{\boldcheck} & \cellcolor{gray!10}\boldcheck & \cellcolor{gray!10}\boldcheck \\
 & ProactiveVA~\cite{proactiveVA2026} \hfill 2026 & \levelfour & \purple{\boldcheck} & \sand{\boldcheck} & \cyan{\boldcheck} & \dawn{\boldcheck} & \blue{\boldcheck} &  & \blue{\boldcheck} & \orange{\boldcheck} & \orange{\boldcheck} & \orange{\boldcheck} & \red{\boldcheck} & \red{\boldcheck} & \red{\boldcheck} & \green{\boldcheck} & \green{\boldcheck} & \green{\boldcheck} & \boldcheck & \boldcheck \\
 & \cellcolor{gray!10}DFAgent~\cite{gantayat2026DfagentNaturalLanguageData} \hfill 2026 & \cellcolor{gray!10}\levelfour & \cellcolor{gray!10}\purple{\boldcheck} & \cellcolor{gray!10}\sand{\boldcheck} & \cellcolor{gray!10} & \cellcolor{gray!10}\dawn{\boldcheck} & \cellcolor{gray!10}\blue{\boldcheck} & \cellcolor{gray!10} & \cellcolor{gray!10}\blue{\boldcheck} & \cellcolor{gray!10}\orange{\boldcheck} & \cellcolor{gray!10}\orange{\boldcheck} & \cellcolor{gray!10} & \cellcolor{gray!10} & \cellcolor{gray!10} & \cellcolor{gray!10} & \cellcolor{gray!10} & \cellcolor{gray!10} & \cellcolor{gray!10} & \cellcolor{gray!10}\boldcheck & \cellcolor{gray!10}\boldcheck \\
 & HiLSVA~\cite{ai2026HilsvaDesignEvaluationHuman} \hfill 2026 & \levelfour & \purple{\boldcheck} & \sand{\boldcheck} & \cyan{\boldcheck} & \dawn{\boldcheck} & \blue{\boldcheck} &  & \blue{\boldcheck} & \orange{\boldcheck} & \orange{\boldcheck} & \orange{\boldcheck} & \red{\boldcheck} &  &  &  &  &  & \boldcheck & \boldcheck \\

      \hline

    \end{tabular}}
\end{table*}

\subsection{Analytical Coding}
\label{sec:analytical-coding}

We coded each primary system along four dimensions selected to connect system architecture to human involvement:
\begin{enumerate}
  \item \textbf{Autonomy level}: the extent to which a system can organize, evaluate, and initiate its own analytical work;
  \item \textbf{Functional role}: whether a system function acts as a \planner, \creator, \reviewer, or \contextmanager;
  \item \textbf{VA workflow stage}: whether the function operates in \dataprocessing, \visualmapping, \viewtransformation, or \presentation; and
  \item \textbf{Human interaction and control}: the responsibilities retained by users, the timing and modality of their interventions, and the mechanisms provided for inspection, correction, approval, and governance.
\end{enumerate}

Autonomy level provides the macro classification of a system. Role~$\times$~Workflow coding provides a finer-grained account of how that autonomy is realized. We treated these dimensions as related but not interchangeable: a system may cover many workflow stages through a fixed pipeline, for example, without satisfying the reflective criterion for Level~3.

\subsection{Operationalizing Autonomy}
\label{sec:level}

We assigned one of four levels using architectural evidence reported in the paper or system documentation. The levels are cumulative in capability but separated by explicit gates:

\begin{description}
  \item[\textbf{Level 1---Reactive Execution.}] The system uses a single model call or a fixed sequence of skills to respond to user input. It neither autonomously decomposes the task nor uses an output-driven evaluation loop to revise its execution.

  \item[\textbf{Level 2---Predefined Workflow.}] The system coordinates multiple stages or specialized functions in a topology fixed at design time. It may clarify requests, maintain session state, or support iterative user correction, but it does not dynamically restructure execution based on intermediate results.

  \item[\textbf{Level 3---Reflective Adaptation.}] The system supports dynamic task decomposition or routing and contains an explicit execute--evaluate--revise loop. A runtime \reviewer{} must assess an intermediate or final artifact and its assessment must be able to change subsequent execution; post-hoc scoring alone does not qualify.

  \item[\textbf{Level 4---Proactive Evolution.}] The system satisfies Level~3 and additionally supports at least one capability that changes when or how work is organized: proactive task initiation, runtime adaptation of workflow topology or agent assignment, or persistent evolution of knowledge, skills, or memory across tasks.
\end{description}

These criteria distinguish autonomy from architectural size. Multiple agents in a static chain remain Level~2, while a single agent may reach Level~3 if it dynamically decomposes work and closes a runtime review loop. Likewise, multi-turn conversation does not by itself constitute reflection: when users identify errors and request revisions, the human rather than the system supplies the reviewing function.

\subsection{Operationalizing Roles and Workflow Stages}
\label{sec:taxonomy_axes}

\paragraph{Functional roles.}
Roles describe functions, not necessarily separately implemented software agents. A component can instantiate more than one role, and several components can share a role.
\begin{itemize}
  \item \planner: clarifies goals, decomposes tasks, selects tools or agents, routes work, or organizes workflow execution.
  \item \creator: executes analytical work by producing transformations, code, visualization specifications, interactive views, annotations, reports, or other artifacts.
  \item \reviewer: evaluates an intermediate or final artifact for properties such as correctness, semantic alignment, visual quality, or cost. We code a runtime reviewing function only when its assessment can inform a later system or human action.
  \item \contextmanager: maintains or retrieves state required across actions, including dialogue history, intermediate representations, data and tool profiles, provenance, and persistent knowledge or skills.
\end{itemize}

\paragraph{Workflow stages.}
We use a four-stage abstraction of the VA workflow to locate where role functions operate:
\begin{itemize}
  \item \dataprocessing: data acquisition, profiling, integration, cleaning, and transformation;
  \item \visualmapping: selection and construction of spatial substrates, graphical elements, and visual encodings;
  \item \viewtransformation: navigation, filtering, hierarchy drilling, and coordination or rearrangement of views; and
  \item \presentation: annotation, summarization, explanation, and storytelling for communicating analytical findings.
\end{itemize}
The \planner{} and \contextmanager{} often span stage boundaries. We therefore code them at every stage whose execution they materially coordinate or support rather than treating planning or memory as additional workflow stages.

\subsection{Coding Procedure and Boundary Cases}
\label{sec:coding-procedure}

Two authors independently screened candidate papers and resolved inclusion disagreements through discussion. During coding, we iteratively refined the codebook against representative systems, recorded evidence for each assigned code, and discussed ambiguous cases until reaching consensus. We assigned the \emph{lowest autonomy level whose required conditions were all explicitly supported}; missing implementation evidence was not interpreted as evidence of a higher-level capability.

Several recurring boundary cases required particular care. We distinguished user-driven iteration from system-driven reflection, static orchestration from runtime workflow adaptation, retrieval from persistent learning, and evaluation benchmarks from deployed runtime \reviewer{} functions. We coded only capabilities that were implemented and described, not capabilities proposed as future work. Because the codebook evolved alongside an emerging and rapidly changing literature, we do not report a formal inter-rater reliability statistic. To support scrutiny and reuse, the companion materials expose system-level codes and the evidence used to resolve these cases.

%!TEX root = ../main.tex

\section{How Agentic Autonomy Reshapes the VA Workflow}
\label{sec:workflow}

Having operationalized autonomy, roles, and workflow stages, we now examine how agent responsibilities change across the VA pipeline as autonomy increases. We organize the findings by \dataprocessing, \visualmapping, \viewtransformation, and \presentation. Within each stage, we trace the observed transition, the corresponding role reconfiguration, representative evidence, remaining human control points, and persistent limitations; each subsection concludes with a situated design implication.

\begin{figure*}[t]
  \centering
    \includegraphics[width=\textwidth]{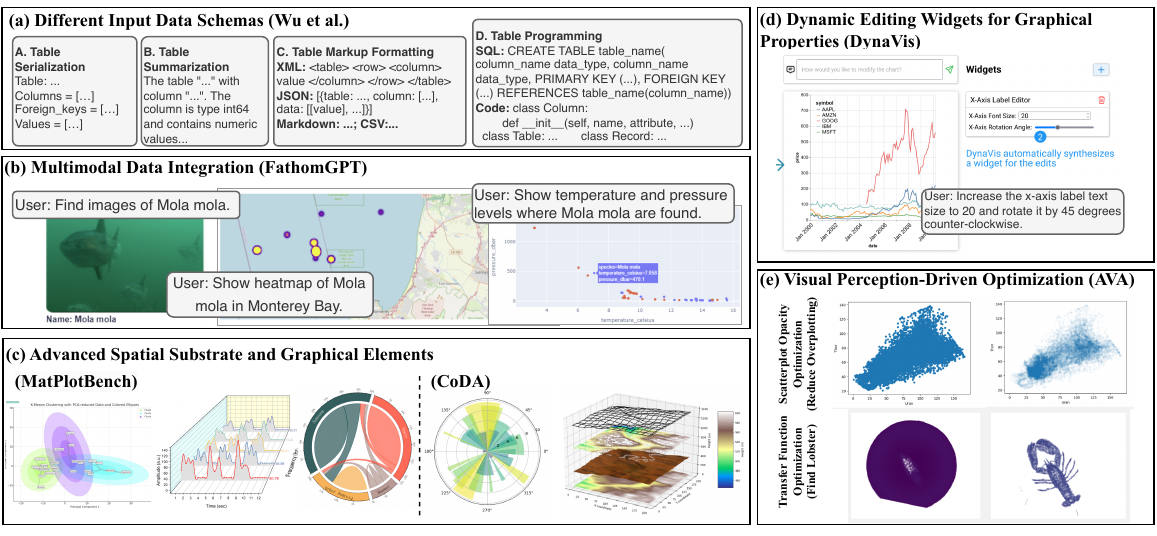}
    \vspace{-2em}
  \caption{Example corpus of paradigm innovations in agentic VA systems. (a) Input Data Schemas: Shifting from static serialization to programming-based representations to enhance LLM reasoning~\cite{wu2024AutomatedDataVisualization}.
(b) Multimodal Data Integration: Incorporating external domain knowledge and unstructured data into analytical pipelines~\cite{khanal2024FathomGPTNaturalLanguage}.
(c) Advanced Structures: Generating high-dimensional spatial substrates and complex graphical elements~\cite{yang2024MatPlotAgentMethodEvaluation, chen2025codaagenticsystemscollaborative}.
(d) Dynamic Editing: Automatically synthesizing UI widgets for fine-grained, intent-driven manipulation of graphical properties~\cite{vaithilingam2024DynaVisDynamicallySynthesized}.
(e) Perception-Driven Optimization: Utilizing MLLM-based reflective loops to perceive and refine visual outputs~\cite{ava2024}.}
  \Description{Five annotated examples adapted from surveyed systems, labeled (a) through (e). They illustrate programming-based input data schemas, multi-modal knowledge integration, generation of advanced visual structures, dynamically synthesized editing widgets, and a perception-driven optimization loop in which a multimodal LLM inspects rendered charts and refines them.}
  \label{fig:example-abc}
\end{figure*}

Two roles operate especially across stage boundaries. The \planner{} coordinates and, at higher autonomy, restructures stage execution; the \contextmanager{} maintains the dialogue, intermediate artifacts, provenance, and knowledge on which that coordination depends. They are therefore not additional workflow stages. Instead, we trace how they support work within each stage below. Fig.~\ref{fig:example-abc} previews representative changes, from data profiling and integration to dynamic visual editing and perception-driven optimization.

\begin{figure*}[t]
  \centering
    \includegraphics[width=\textwidth]{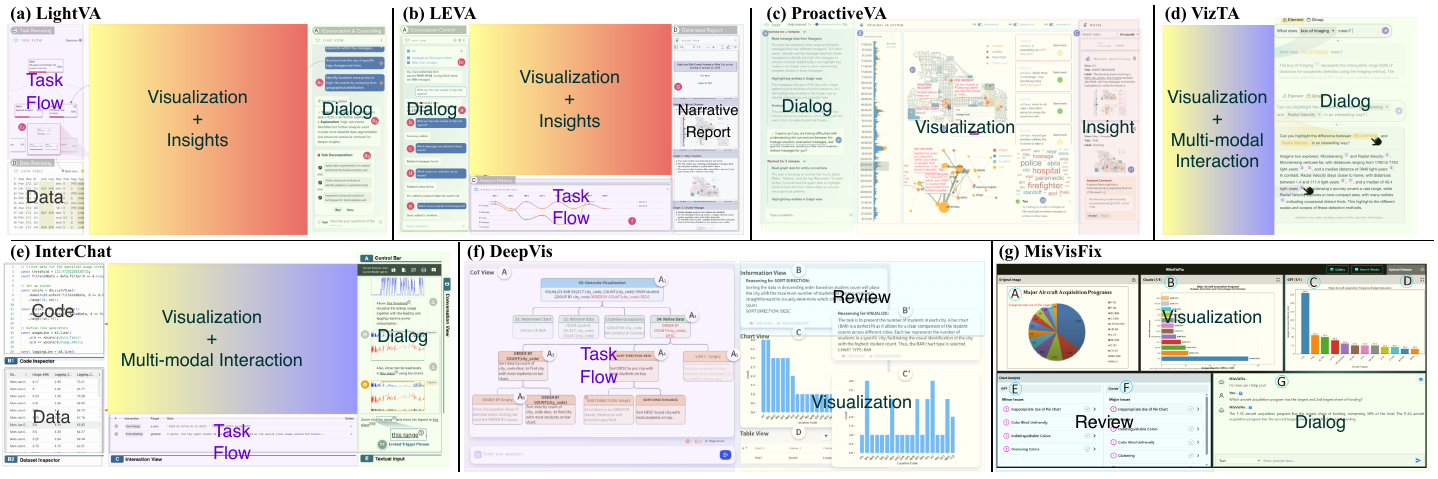}
    \vspace{-2em}
  \caption{Representative Human-AI Interaction Interfaces in Agentic Visual Analytics.
\textbf{Task Workflow:}
(a), (f) provide interactive task flows allowing users to directly manipulate, approve, or modify agentic decomposition plans.
(b), (e) maintain temporal creation histories, enabling users to conceptualize the analytical process and track session evolution.
\textbf{Narrative Data Facts:}
In addition, (a), (b) and (c) highlight the generation of summarized data facts and structured narrative reports to bridge the semantic gap.
\textbf{Multi-modal Interaction:}
(d), (e) combine natural language dialogue with direct spatial selections (e.g., lassoing) to ground the agent's reasoning in the user's visual focus.
\textbf{Structured Visual Reviews:}
(f), (g) displays the review details, exposing internal evaluation logic and discrepancies for iterative visual correction.}
  \Description{Seven interface screenshots adapted from surveyed systems, labeled (a) through (g). They show interactive task-flow editors for approving or modifying agent plans, timelines of session history, panels summarizing narrative data facts, multi-modal input combining chat with lasso selections on charts, and structured review panels that expose the agent's evaluation logic for iterative correction.}
  \label{fig:example-interface}
\end{figure*}

\subsection{Data Processing: From Static Schema to Autonomous Exploration}

\textbf{Traditional \blue{data processing} in visual analytics transforms raw data into visualization-ready structures, typically involving acquisition, profiling, integration, and transformation.} Historically, this stage relied heavily on structured data with predefined schemas. Early rule-based systems like Articulate~\cite{tabalba2022ArticulateAlwaysListeningNatural} and even initial LLM-based tools like ncNet~\cite{luo2022NaturalLanguageVisualization} and Chat2VIS~\cite{maddigan2023Chat2VISGeneratingData} were highly sensitive to schema variations, requiring manual or semi-automatic interventions for profiling and integration based on preset rules. In our corpus, \dataprocessing is the second-most covered stage (61 of 65 systems), and agentic capabilities within it cluster into three groups.

\textbf{The first cluster is \blue{schema and relational reasoning}.} Agents no longer rely on fixed data schemas: as shown in Fig.~\ref{fig:example-abc}(a), the shift from static table serialization to programming-based schema representations, such as those proposed by Wu et al.~\cite{wu2024AutomatedDataVisualization}, significantly enhances an LLM's in-context reasoning. Systems like Prompt4Vis~\cite{li2024Prompt4VisPromptingLarge} utilize schema linking for multi-table integration, while PI2~\cite{chen2022PI2EndtoendInteractive} ensures transparent integration by presenting the intermediate joined tables to users. Leveraging these representations, agents perform dynamic entity recognition, relationship extraction, and concept alignment on the fly rather than depending on manually matched join keys.

\textbf{The second cluster is \blue{generalized data modalities}.} Agents process sources of increasing complexity, moving beyond tabular data to unstructured documents, images, and domain-specific formats. As shown in Fig.~\ref{fig:example-abc}(b), FathomGPT~\cite{khanal2024FathomGPTNaturalLanguage} integrates multi-modal external domain knowledge into its analytical pipeline; LIDA~\cite{dibia2023LIDAToolAutomatic} processes multiple data formats to generate comprehensive data summaries; and Data Formulator~\cite{wang2023DataFormulatorAIpowered} specializes in transforming complex, non-standard formats. Recent systems push into specialized domains: SUPQA~\cite{huang2025SUPQALLMbasedGeoVisualization} and GeoAuthor~\cite{chen2026GeoauthorLinkingTextVisualization} handle geographic data, PhenoFlow~\cite{kim2024PhenoFlowHumanLLMDriven} wrangles large-scale, irregular clinical records, and HiLSVA~\cite{ai2026HilsvaDesignEvaluationHuman} and SASAV~\cite{sun2026SasavSelfDirectedAgent} extend agentic processing to volumetric scientific data.

\textbf{The third cluster is \blue{autonomous profiling and metadata generation}, where the division of labor among roles is most visible.} CoDA~\cite{chen2025codaagenticsystemscollaborative} performs profiling and identifies necessary transformations without loading the full dataset into the LLM context, so that the \planner and \reviewer reason over structural summaries while the \creator emits transformation code. LightVA~\cite{zhao2025LightVALightweightVisual} generates a ``Data Introduction''---a JSON structure enriched with natural language descriptions and statistics---to bridge the semantic gap between raw data and the LLM's reasoning. Gyarmati et al.~\cite{gyarmati2025composableagenticautomatedvisual} push this further with a field expander equipped with web search tools to resolve cryptic codes, alongside a dataset profiler that externalizes ``statistical truth'' so the agent does not hallucinate data distributions. Transformation itself becomes intent-driven: Data Formulator~2~\cite{wang2025DataFormulator2} and DataWeaver~\cite{fu2025DATAWEAVERAuthoringDataDriven} automatically apply binning strategies or add trendlines based on inferred analytical intent, and ProactiveVA~\cite{proactiveVA2026} proactively suggests filtering out specific data subsets (e.g., negative profit ratios) that align with the user's implicit direction of exploration.

\textbf{Human control points in this stage concentrate on \blue{gatekeeping and verification}.} PhenoFlow executes all generated wrangling code locally, keeping raw patient data out of external APIs and leaving clinicians in charge of what leaves the premises; profiling outputs are surfaced for inspection (PI2's intermediate tables, LightVA's Data Introduction) precisely because a wrong profile silently corrupts every downstream stage.

\nistart{Discussion.}
\textbf{Despite these advancements, several bottlenecks remain in agentic \blue{data processing}.} The most prominent is the LLM context window limit: although the \contextmanager mitigates the problem, maintaining a complete profile and processing logic for ultra-large datasets over long sessions remains difficult, as observed in AI Threads~\cite{hong2023ConversationalAIThreads}. Agents also face a severe semantic gap in heterogeneous data lakes lacking unified metadata. Finally, profiling itself can hallucinate---the very motivation for Gyarmati et al.'s externalized statistics---and autonomous exploration raises security and privacy risks when agents inadvertently pass sensitive values to external APIs.

\implicationbox{DP-1}{\textbf{Isolate raw data from the LLM context.} For systems at Level~2 and above, and especially for users in privacy-sensitive domains, let the \planner and \reviewer operate on structured data profiles while deterministic transformations are routed to generated code executed outside the model. This pattern is explicit in CoDA~\cite{chen2025codaagenticsystemscollaborative}, PhenoFlow~\cite{kim2024PhenoFlowHumanLLMDriven}, and LightVA~\cite{zhao2025LightVALightweightVisual}, and implicit in many more systems. \emph{Boundary:} profiles can themselves be hallucinated---pair them with externally computed statistics as in Gyarmati et al.~\cite{gyarmati2025composableagenticautomatedvisual}.}

\subsection{Visual Mapping: From Simple Structures to Feature-Rich Encodings}

\textbf{\orange{Visual mapping}, the core of visual analytics, transforms abstract data into perceivable forms through spatial substrate and graphical element selection.} Traditional natural language interfaces like NL4DV~\cite{narechania2021NL4DVToolkitGenerating} primarily map queries to predefined grammars, relying on explicit encoding specifications and relatively simple chart structures. \visualmapping is the most covered stage in our corpus (63 of 65 systems), and its agentic capabilities divide naturally into what agents contribute and what humans retain.

\begin{figure*}[t]
\centering
 \begin{minipage}[t]{0.49\linewidth}
   \centering
   \includegraphics[width=\linewidth]{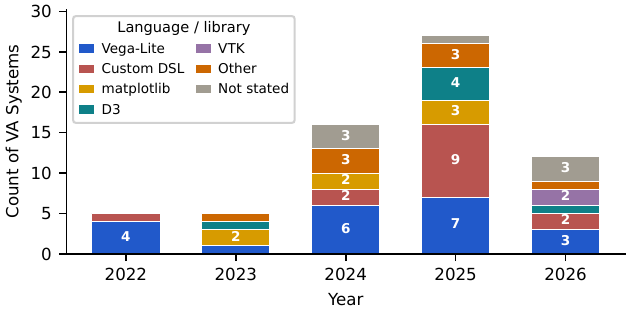}\\[1pt]
   {\small (a)}
 \end{minipage}\hfill
 \begin{minipage}[t]{0.49\linewidth}
   \centering
   \includegraphics[width=\linewidth]{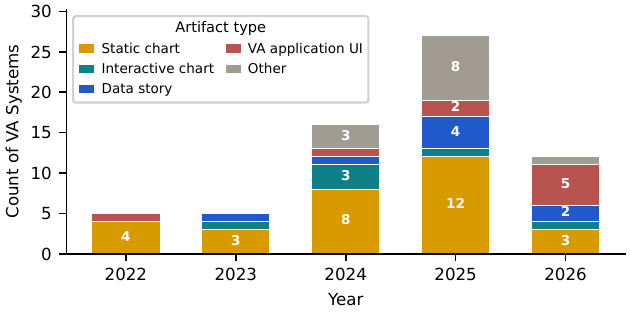}\\[1pt]
   {\small (b)}
 \end{minipage}
\caption{\creator{} output trends across the 65 surveyed systems. (a) Primary generation language or library per year. (b) Primary artifact type per year.}
\Description{Two side-by-side stacked bar charts. Chart (a) shows the number of systems per year by primary generation language or library: Vega-Lite remains dominant across years, custom domain-specific languages surge in 2025, matplotlib declines after 2024, and VTK first appears in 2026. Chart (b) shows systems per year by artifact type: static charts dominate but their share declines, while full VA application interfaces rise sharply in 2026 and data stories appear steadily from 2025.}
\label{fig:mapping-trends}
\vspace{-1em}
\end{figure*}

\textbf{On the AI side, the first capability is \orange{complex structure mapping and multi-library generation}.} As shown in Fig.~\ref{fig:example-abc}(c), systems like MatPlotAgent~\cite{yang2024MatPlotAgentMethodEvaluation} and CoDA~\cite{chen2025codaagenticsystemscollaborative} generate advanced spatial substrates and diverse graphical elements far beyond basic chart types, while LIDA~\cite{dibia2023LIDAToolAutomatic} switches seamlessly between Matplotlib, Plotly, and Altair. The second is \orange{\textbf{example- and rule-grounded generation}}. Level~1 systems such as RGVisNet~\cite{song2022RGVisNetHybridRetrievalGeneration} pioneered retrieval-guided generation over static code corpora; agentic systems make the grounding dynamic and open-ended. CoDA retrieves real-world code snippets from online galleries, Shao et al.~\cite{shao2025dolanguagemodelagents} add a retrieval-augmented \reviewer{} that validates the \creator's reasoning against externally sourced design principles, and Gyarmati et al.~\cite{gyarmati2025composableagenticautomatedvisual} prime Draco~\cite{2019-draco}, a deterministic rule engine, with LLM-generated metadata so that principled design choices are not left to the language model. The third is the \orange{\textbf{rendered-output review loop}} powered by multi-modal LLMs: agentic systems map visual structures back to analytical actions. MisVisFix~\cite{misvisfix2026} detects and corrects misleading encodings such as truncated axes, surfacing its structured review in the interface (Fig.~\ref{fig:example-interface}(g)); AVA~\cite{ava2024} evaluates open objectives such as ``does this show the structure of interest?'' (Fig.~\ref{fig:example-abc}(e)); and $C^2$~\cite{koh2025C^2ScalableAutoFeedback} scales such feedback for chart generation.

\textbf{On the human side, agents widen rather than replace user control.} Users express \orange{\textbf{open-ended intent}}: Chat2VIS~\cite{maddigan2023Chat2VISGeneratingData} and Text2Chart31~\cite{pesaranzadeh2024Text2Chart31InstructionTuning} interpret deep analytical goals through dialog, and LightVA~\cite{zhao2025LightVALightweightVisual} maps abstract objectives (``identify high-risk events'') to concrete visual types---a flexibility beyond earlier task-based recommenders like TaskVis~\cite{taskvis2021shen}. Outside the LLM setting, HAIChart~\cite{xie2024haichart} makes this pairing explicit: it first proposes charts, then folds user-selected visualization hints back into the search so later recommendations follow evolving preferences. Users also ground intent \orange{\textbf{by example and by space}}: InterChat~\cite{chen2025InterChatEnhancingGenerative} accepts sketched patterns and spatial selections, and Data Formulator~\cite{wang2023DataFormulatorAIpowered} anchors encodings in drag-and-drop field mappings. Finally, users intervene \orange{\textbf{surgically without full regeneration}}: DynaVis~\cite{vaithilingam2024DynaVisDynamicallySynthesized} synthesizes UI widgets on demand for fine-grained property edits (Fig.~\ref{fig:example-abc}(d)), NL4DV-Stylist~\cite{ji2025NL4DVStylistStylingData} and Blace~\cite{lyi2024LearnableExpressiveVisualization} target styling, and DataWeaver~\cite{fu2025DATAWEAVERAuthoringDataDriven} lets edits propagate without discarding the surrounding artifact.

\textbf{Two corpus-level trends sharpen this picture} (Fig.~\ref{fig:mapping-trends}). First, in generation targets, Vega-Lite remains the stable backbone (21 of 65 systems), but custom domain-specific languages surge to nine systems in 2025 alone---Draco priming in Gyarmati et al., the visualization query language of Prompt4Vis~\cite{li2024Prompt4VisPromptingLarge}, and BONSAI's constrained grammar~\cite{spinner2026BonsaiMixedInitiativeWorkspace}---while matplotlib recedes from its 2023--2024 peak and VTK first appears in 2026 with scientific visualization systems~\cite{ai2026HilsvaDesignEvaluationHuman,sun2026SasavSelfDirectedAgent}. The \creator's output is shifting from direct plotting code toward \orange{\textbf{constrained intermediate representations}}, which couple naturally with the \reviewer's need for verifiable artifacts. Second, in artifact granularity, static charts still dominate (30 of 65) but their share declines: full VA application interfaces reach five of twelve 2026 systems (BONSAI, HiLSVA, DFAgent~\cite{gantayat2026DfagentNaturalLanguageData}, ProactiveVA~\cite{proactiveVA2026}), and data stories appear steadily from 2025. Agents increasingly build or drive \orange{\textbf{whole applications and narratives}} rather than single images.

\textbf{The effectiveness of the \reviewer{} depends critically on its \orange{information channel}.} ChartEdit~\cite{zhao-etal-2025-chartedit} and MatPlotAgent~\cite{yang2024MatPlotAgentMethodEvaluation} empirically show that reviewers with access to the rendered chart (and, in ChartEdit, the generating code) outperform text-only reviewers. Benchmarks delimit the channels from the other side: ChartEditVista~\cite{chen2026CharteditorReinforcementLearningFramework} shows that editing from the chart image alone---without source code---remains markedly harder, and Dashboard2Code~\cite{niu2026Dashboard2codeEvaluatingMultimodalModels} requires agents to interactively probe a live dashboard, since a static screenshot cannot reveal its interaction logic.

\nistart{Discussion.}
\textbf{Despite the integration of MLLMs, agents still suffer from a limited understanding of \orange{visual semantics and aesthetics}.} 
As demonstrated by the VisEval benchmark~\cite{chen2025VisEvalBenchmarkData}, MLLMs like GPT-4o often struggle to accurately interpret and revise chart layouts based on subtle visual cues, frequently misinterpreting text annotations or failing to adjust to a human's desired level of detail. VisJudge-Bench~\cite{xie2025visjudge} isolates the judgment problem: even strong MLLMs diverge from expert ratings of data fidelity, expressiveness, and aesthetics across single charts, multi-chart views, and dashboards. 
Semantic drift in high-dimensional mapping remains a persistent issue, and agents still struggle to capture idiosyncratic user styles and domain-specific preferences.

\implicationbox{VM-1}{\textbf{Give the \reviewer{} both the rendered output and the generating code.} Designers of Level~3 reflective loops should feed the \reviewer{} a dual channel---raster image plus source specification---rather than either alone. Supported by ChartEdit~\cite{zhao-etal-2025-chartedit}, MatPlotAgent~\cite{yang2024MatPlotAgentMethodEvaluation}, and AVA~\cite{ava2024}. \emph{Boundary:} when only images are available, expect a substantially lower performance ceiling, as quantified by ChartEditVista~\cite{chen2026CharteditorReinforcementLearningFramework}.}

\implicationbox{VM-2}{\textbf{Route deterministic design decisions to rules, not prompts.} Where design knowledge is formalizable, let a rule engine or compiler decide and reserve the LLM for semantic gap-filling, as in Gyarmati et al.'s Draco priming~\cite{gyarmati2025composableagenticautomatedvisual} and LEVA's rule-based computation~\cite{zhao2024LEVAUsingLarge}. \emph{Boundary:} this trades open-endedness for reliability and suits well-charted design subspaces.}

\subsection{View Transformation: Complex and High-Dimensional Navigation}

\begin{figure}[t]
\centering
 \includegraphics[width=\columnwidth]{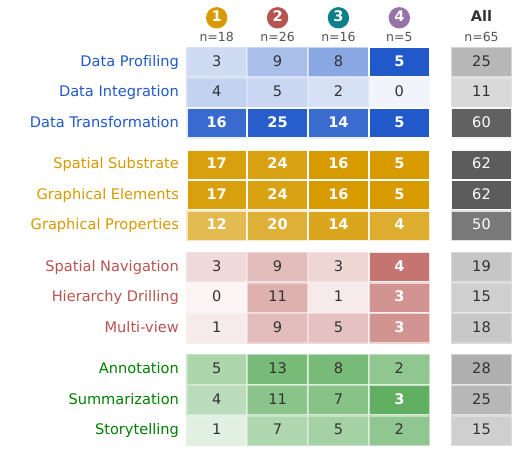}
 \vspace{-0.5em}
\caption{Corpus coverage of the twelve fine-grained workflow capabilities (rows, grouped and colored by stage as in Table~\ref{tab:survey_papers_final}) across autonomy levels (columns); cell shading is normalized by the number of systems at each level. \viewtransformation{} is the thinnest stage and its coverage concentrates at Level~2---hierarchy drilling nearly disappears among Level~3 systems---whereas \presentation{} coverage deepens as autonomy rises.}
\Description{A heatmap with twelve rows for fine-grained workflow capabilities, grouped into data processing, visual mapping, view transformation, and presentation, and four columns for autonomy levels one through four plus a total column. Visual mapping and data transformation rows are covered by nearly all systems at every level. The three view transformation rows are sparsely covered, with hierarchy drilling supported by eleven Level 2 systems but only one Level 3 system. Presentation rows show moderate coverage that grows with autonomy level.}
\label{fig:stage-level}
\vspace{-1em}
\end{figure}

\textbf{\red{View transformation} encompasses the interactions users employ to explore data and shift perspectives:} spatial navigation for granularity changes, hierarchy drilling, multi-view coordination (e.g., linking and brushing), and filtering to modify displayed data subsets. This is by far the thinnest stage in our corpus---only 27 of 65 systems (42\%) touch it (Fig.~\ref{fig:stage-level})---so the patterns below should be read as emerging rather than established. Coverage is also uneven across levels: hierarchy drilling concentrates in Level~2 systems and nearly vanishes at Level~3, suggesting that reflective systems have so far prioritized generation over interactive navigation. Within this smaller evidence base, agentic capabilities organize into five recurring themes.

\textbf{The first theme is \red{interaction state and provenance}.} LEVA~\cite{zhao2024LEVAUsingLarge} and InterChat~\cite{chen2025InterChatEnhancingGenerative} maintain temporal creation histories that let users conceptualize the analytical process (Fig.~\ref{fig:example-interface}(b),(e)), and FlowForge~\cite{flowforge2026} persists a design space of explored alternatives so users can pivot among workflow versions; AI Threads~\cite{hong2023ConversationalAIThreads} illustrates the negative case, where state drifts across long conversations.

\textbf{The second theme is \red{spatial navigation and hierarchical drilling}.} Intelligent Drill-Down~\cite{zheng2026IntelligentDrillDownLarge} infers analytical intent from the user's interaction trajectory and recommends drill-down paths while leaving the choice to the analyst---a division of labor between behavioral inference and human selection. Plume~\cite{lisnic2025PlumeScaffoldingText} uses frame dependency trees for progressive navigation of nested relationships, SUPQA~\cite{huang2025SUPQALLMbasedGeoVisualization} and FathomGPT~\cite{khanal2024FathomGPTNaturalLanguage} navigate domain spaces, and HiLSVA~\cite{ai2026HilsvaDesignEvaluationHuman} and SASAV~\cite{sun2026SasavSelfDirectedAgent} extend navigation to 3D camera control over volumetric data.

\textbf{The third theme is \red{multi-view layout, coordination, and conflict}.} As shown in Fig.~\ref{fig:example-interface}(a), LightVA~\cite{zhao2025LightVALightweightVisual} merges user-selected tasks into a coordinated dashboard with shared brushes; PhenoFlow~\cite{kim2024PhenoFlowHumanLLMDriven} coordinates cohort-transition and patient-overview views; Aggarwal et al.~\cite{aggarwal2025GoalDrivenDataStory} adjust sizing and positioning as narrative requirements evolve; and BONSAI~\cite{spinner2026BonsaiMixedInitiativeWorkspace} raises the granularity to application-level multi-view workspaces co-constructed with the user.

\textbf{The fourth theme is \red{grounding dialog in direct manipulation}.} InterChat converts lasso and box selections into textual descriptors that the agent uses to reconfigure or zoom into views (Fig.~\ref{fig:example-interface}(e)); VizTA~\cite{vizTA2025} makes references in its dialog clickable, linking each explanation back to the visual element it describes; and RECITKIT~\cite{setlur2025RECITKITSpatialToolkit} extends spatial grounding to immersive environments.

\textbf{The fifth theme is \red{agent-initiated reconfiguration}---and its cost.} ProactiveVA~\cite{proactiveVA2026} can ``take over'' the workflow to perform multi-view synchronization (e.g., clicking a hexagon in a map to update a corresponding message view) based purely on its internal reasoning (Fig.~\ref{fig:example-interface}(c)), and DASH~\cite{bromley2024DASHBimodalData} restructures views around drag-and-drop context. These are the strongest form of agency in this stage, and also the most disorienting when the interface changes in ways the user did not anticipate.

\nistart{Discussion.}
\textbf{The primary bottleneck is complex interactive \red{state management}:} across extensive multi-turn dialog, accurately maintaining interaction states and historical context remains challenging, as evidenced by AI Threads~\cite{hong2023ConversationalAIThreads}. Conflict resolution in multi-view coordination is similarly unsolved---synchronizing interconnected views can produce unpredictable UI behavior. Above all, agent-initiated view changes shift cognitive load rather than remove it: the user saves manipulation effort but must re-orient after every reconfiguration they did not perform themselves.

\implicationbox{VT-1}{\textbf{Make agent-initiated view changes reversible and legible.} Level~3 and Level~4 multi-view systems should pair every agent-initiated reconfiguration with an undo path and an explicit indication of what changed, following the takeover affordances of ProactiveVA~\cite{proactiveVA2026} and the task-flow visibility of LightVA~\cite{zhao2025LightVALightweightVisual}. \emph{Boundary:} with only 27 systems covering this stage, we state this as a preliminary pattern rather than an established rule.}

\subsection{Presentation: From Templates to Semantic Storytelling}

\textbf{The \green{presentation} stage communicates analytical findings through summarization, annotation, and storytelling.} Historically, it relied on rigid templates and static images, requiring users to manually craft narratives around exported charts, as in early systems like DataTales~\cite{sultanum2023DATATALESInvestigatingUse}. In our corpus, 36 of 65 systems cover this stage, with coverage deepening as autonomy rises (Fig.~\ref{fig:stage-level}), and their outputs serve two distinct purposes: communicating finished findings to an audience, and supporting the user's own ongoing exploration.

\textbf{For \green{communication-oriented} outputs, agents autonomously assemble structured, coherent artifacts.} LightVA~\cite{zhao2025LightVALightweightVisual} synthesizes findings from multiple subtasks into a cohesive narrative, with the \planner and \contextmanager enforcing visual consistency constraints (color, layout) across the document. Gyarmati et al.~\cite{gyarmati2025composableagenticautomatedvisual} pair a vision-language report narrator with a dual-output interface---an interactive Observable report for the reader and an executable notebook for the analyst---so the user can surgically debug a specific agent's logic without regenerating the whole report. DataNarrative~\cite{islam2024DataNarrativeAutomatedDataDriven} generates multi-modal presentations, Jupybara~\cite{wang2025JupybaraOperationalizingDesign} operationalizes the computational-notebook paradigm for iterative storytelling, and Aggarwal et al.~\cite{aggarwal2025GoalDrivenDataStory} structure narratives around evolving analytical goals. Two recent systems foreground \green{\textbf{authorial control}}: Vistoryteller~\cite{shi2026VistorytellerDesigningDataStories} organizes data-story production as a newsroom-style division of agent labor while giving the author fine-grained control over both local wording and global narrative arc, and GeoAuthor~\cite{chen2026GeoauthorLinkingTextVisualization} maintains bidirectional consistency between article text and linked maps, so edits on either side propagate to the other. After our August 2026 cutoff, DataMagic~\cite{xie2026datamagic} extends this stage from static stories to narrated data videos: agents generate candidate scenes in parallel, a later orchestration pass orders them for narrative coherence, and a declarative specification keeps charts, narration, and animation bound to the underlying data while remaining editable.

\textbf{For \green{exploration- and education-oriented} outputs, agents annotate and explain in place.} LEVA~\cite{zhao2024LEVAUsingLarge} overlays intelligent annotations onto existing VA systems and compiles paginated LaTeX reports combining screenshots with textual summaries (Fig.~\ref{fig:example-interface}(b)); DataWeaver~\cite{fu2025DATAWEAVERAuthoringDataDriven} anchors generated ``data facts'' directly onto charts as layered overlays; PhenoFlow~\cite{kim2024PhenoFlowHumanLLMDriven} pairs visual output with plain-language explanations; VizTA~\cite{vizTA2025} explains chart semantics grounded in statistical definitions; and ProactiveVA~\cite{proactiveVA2026} pushes note cards highlighting critical details as the user explores.

\textbf{Across both purposes, the analytically load-bearing property is \green{traceability from claim to evidence}.} DataWeaver's anchored data facts and GeoAuthor's text--visualization links keep every narrative statement attached to the data or code that produced it, and Gyarmati et al.'s executable notebook makes the full derivation inspectable. Where such anchoring is absent, readers must take the agent's narrative on faith---precisely the failure mode that misinformation-oriented work like MisVisFix~\cite{misvisfix2026} targets from the reviewing side.

\nistart{Discussion.}
\textbf{Maintaining long-range \green{narrative coherence} remains difficult:} agents struggle to preserve global logical consistency in long reports due to context window limits. Assessing the novelty and value of insights is inherently hard, as agents often highlight statistically significant but practically irrelevant findings. Finally, ethics and bias propagation present ongoing challenges, since generated narratives can amplify underlying data biases while sounding authoritative.

\implicationbox{PR-1}{\textbf{Anchor every narrative claim to its underlying artifact.} Storytelling systems at Level~2 and above serving authors should link each generated claim to the data subset, chart, or code cell that supports it, as in DataWeaver~\cite{fu2025DATAWEAVERAuthoringDataDriven}, GeoAuthor~\cite{chen2026GeoauthorLinkingTextVisualization}, and LEVA~\cite{zhao2024LEVAUsingLarge}. \emph{Boundary:} current anchoring is mostly one-directional; bidirectional text--visualization consistency has a single exemplar (GeoAuthor) in the corpus.}

%!TEX root = ../main.tex

\section{How Agentic Autonomy Reshapes Human Work}
\label{sec:interaction}

The workflow changes documented in Section~\ref{sec:workflow} alter what people must do, when they must intervene, and what they must be able to inspect. We examine this redistribution along three dimensions: human responsibilities and burdens across autonomy levels, cooperation across time, and control modalities for different users and situations. Fig.~\ref{fig:example-interface} illustrates the resulting design space, including editable task plans, persistent histories, multimodal grounding, and structured reviews. Across these dimensions, autonomy reduces direct operation but creates new work in verification, permission setting, provenance checking, and governance.

\subsection{Human Roles, Expertise, and Burden Across Autonomy Levels}
\label{sec:human-roles}

Agentic systems demand not one kind of user expertise but four, in varying mixtures: domain knowledge, visualization literacy, tool and programming skill, and---increasingly---\emph{AI supervision skill}, the ability to read an agent's plans and evidence and to judge when to intervene. Examining our corpus level by level shows how the human's responsibilities, and the expertise they draw on, are redistributed rather than simply reduced.

At Level~1 (18 systems), humans do everything but the synthesis itself: they specify complete queries, resolve ambiguity, and review every output. Lu et al.~\cite{lu2026RobustInteractiveTextVisualization} mark the lower boundary of human-agent division of labor with a single clarify-and-confirm turn for ambiguous queries. At Level~2 (26 systems), humans direct a predefined process: they approve clarifications and control handoffs between fixed stages. Visualization Autocomplete~\cite{jeon2026VisualizationAutocompleteVisualizationAuthoring} makes the delegation granularity itself adjustable, letting authors slide between manual specification and agent completion, while InterChat~\cite{chen2025InterChatEnhancingGenerative} lets users steer each round through combined dialog and manipulation.

At Level~3 (16 systems), the human's object of work changes from outputs to \emph{plans and evidence}. Users approve or edit decomposition plans---LightVA~\cite{zhao2025LightVALightweightVisual} exposes a task flow where an AND operator can be changed to a DOWN operator (Fig.~\ref{fig:example-interface}(a)), CoDA~\cite{chen2025codaagenticsystemscollaborative} publishes a global TODO list, and DeepVis~\cite{shuai2025DeepVISBridgingNatural} makes its chain of reasoning inspectable (Fig.~\ref{fig:example-interface}(f)). Because Level~3 systems bring their own AI \reviewer{} (16 of 16 in our corpus), the human increasingly audits the \emph{reviewer} rather than the artifact: Shao et al.~\cite{shao2025dolanguagemodelagents} find that agent evaluation is most reliable when anchored by high-confidence expert hypotheses, and PhenoFlow~\cite{kim2024PhenoFlowHumanLLMDriven} and InterChat expose intermediate data and code so a human reviewer can formulate targeted corrections. At Level~4 (5 systems), humans set goals, permissions, and policies, and teach preferences: FlowForge~\cite{flowforge2026} elevates the user to an architect of the agent workflow itself, ProactiveVA~\cite{proactiveVA2026} lets users tune when behavioral monitoring may intervene, HiLSVA~\cite{ai2026HilsvaDesignEvaluationHuman} routes agent plans through human approval and learns from the feedback, and DFAgent~\cite{gantayat2026DfagentNaturalLanguageData} makes users curators of a growing tool catalog distilled from their own past interactions.

Two burdens move in opposite directions across this progression (Fig.~\ref{fig:teaser}). The \emph{operational} burden---writing queries, wrangling data, assembling charts---falls monotonically as execution is delegated. The \emph{verification and governance} burden does not: it is high but local at Level~1 (check each output), moderate at Level~2 (stage-wise checks), then rises again at Level~3, where users must understand agent evidence and failure modes, and peaks at Level~4, where accountability, permissions, and cross-session memory must be actively governed. Rising autonomy converts labor from doing to overseeing; it does not abolish it.

Two corpus systems show that even ``overseeing'' understates the variety of human positions. Tytarenko et al.~\cite{tytarenko2026ConfirmationBiasAwarenessVisual} invert the supervision direction entirely: their agent monitors the \emph{human's} interaction traces for confirmation bias and warns the analyst, while the human retains epistemic authority over conclusions. BONSAI~\cite{spinner2026BonsaiMixedInitiativeWorkspace} positions the human as a peer co-builder of a shared workspace rather than a supervisor above it. The trajectory is therefore best described not as ``operator becomes supervisor'' but as a \emph{redistribution of responsibility} whose direction depends on task, risk, and expertise.

\implicationbox{IR-1}{\textbf{Design reversible handoffs with risk- and expertise-aware checkpoints.} Autonomy should be adjustable per task rather than fixed per system, and every delegation should surface an inspectable, interruptible plan with evidence---not a raw chain of thought. Supported by LightVA's editable task flows~\cite{zhao2025LightVALightweightVisual}, CoDA's global TODO list~\cite{chen2025codaagenticsystemscollaborative}, and HiLSVA's plan-approval loop~\cite{ai2026HilsvaDesignEvaluationHuman}. \emph{Boundary:} human attention is a scarce channel; at Level~4, transparency must be layered (summary first, evidence on demand) or it becomes its own burden.}

\subsection{Temporal Dynamics: Cooperation Across Three Timescales}

Traditional natural language interfaces are one-off: the user specifies a complete request in a single prompt. Agentic cooperation instead unfolds on three nested timescales---within a turn, within a session, and across sessions---each with its own state, failure modes, and human role.

\begin{itemize}
  \item \emph{Within a turn (seconds).}
 Systems close an execute--evaluate--revise loop before the user sees any output: the \creator{} generates, the \reviewer{} evaluates, and revision continues until a stopping criterion is met. This is the defining mechanism of Level~3---all 16 systems at that level include a runtime \reviewer. MatPlotAgent~\cite{yang2024MatPlotAgentMethodEvaluation} iterates against rendered output, $C^2$~\cite{koh2025C^2ScalableAutoFeedback} scales automatic visual feedback, and SASAV~\cite{sun2026SasavSelfDirectedAgent} employs an MLLM judge over volume renderings. Characteristic failure modes are erroneous reviews and non-convergent loops; the human's role is to calibrate the loop---setting stopping criteria and deciding which failures escalate.

  \item \emph{Within a session (minutes to hours).}
 The challenge is continuity: preserving and exposing evolving shared state so users can build incrementally. ChartEdit~\cite{zhao-etal-2025-chartedit} accepts ``delta'' instructions (``remove the color bar'') against a shared analytical state; LEVA~\cite{zhao2024LEVAUsingLarge} and InterChat~\cite{chen2025InterChatEnhancingGenerative} preserve full step histories that users can revisit and branch from (Fig.~\ref{fig:example-interface}(b),(e)); and MMCoVisNet~\cite{song2024MarryingDialogueSystems} grounds each turn in the dialog history. Continuity also demands \emph{semantic coupling} between the dialog and the views it manipulates: in most systems the mapping between a request and its visual consequence stays implicit, whereas VizTA~\cite{vizTA2025} embeds clickable references in the dialog that highlight the corresponding visual elements (Fig.~\ref{fig:example-interface}(d)). The characteristic failure mode is state drift---AI Threads~\cite{hong2023ConversationalAIThreads} documents how long conversations silently lose earlier constraints as context is truncated or compressed---and the human's role is to inspect, roll back, or branch from the shared state.

  \item \emph{Across sessions (days to months).}
 The corpus distinguishes several qualitatively different memory substrates that are often conflated. Retrieval-augmented external memory (Prompt4Vis~\cite{li2024Prompt4VisPromptingLarge}, FathomGPT~\cite{khanal2024FathomGPTNaturalLanguage}) recalls examples and domain knowledge; executable artifacts (Gyarmati et al.'s notebooks~\cite{gyarmati2025composableagenticautomatedvisual}) persist the full derivation for later re-execution; skill and tool catalogs (DFAgent~\cite{gantayat2026DfagentNaturalLanguageData}) turn past interactions into reusable capabilities; and learned adaptation updates the agent itself, as in HiLSVA's test-time knowledge base~\cite{ai2026HilsvaDesignEvaluationHuman} and the strategy-teaching interface of MisVisFix~\cite{misvisfix2026}. Persistent preference learning and reinforcement-based personalization remain aspirational---no system in our corpus demonstrates them with peer-reviewed evidence. Characteristic failure modes are stale, incorrect, or uncontrollable persistence; the human's role is to inspect, retract, and govern what the agent retains---yet only five corpus systems persist state across sessions at all.
\end{itemize}

\implicationbox{IR-2}{\textbf{Scope state explicitly, and give users control over what persists.} Systems should separate turn-level, session-level, and cross-session state; attach provenance to each remembered item; and let users inspect and retract what the agent retains, as in CoDA's shared task state~\cite{chen2025codaagenticsystemscollaborative}, FlowForge's persistent design space~\cite{flowforge2026}, and HiLSVA's user-approved knowledge base~\cite{ai2026HilsvaDesignEvaluationHuman}. \emph{Boundary:} cross-session memory governance has no mature exemplar---only five corpus systems persist state across sessions at all.}

\subsection{Modalities of Control: Matching Channels to Users and Scenarios}

\begin{figure*}[t]
  \centering
    \includegraphics[width=\textwidth]{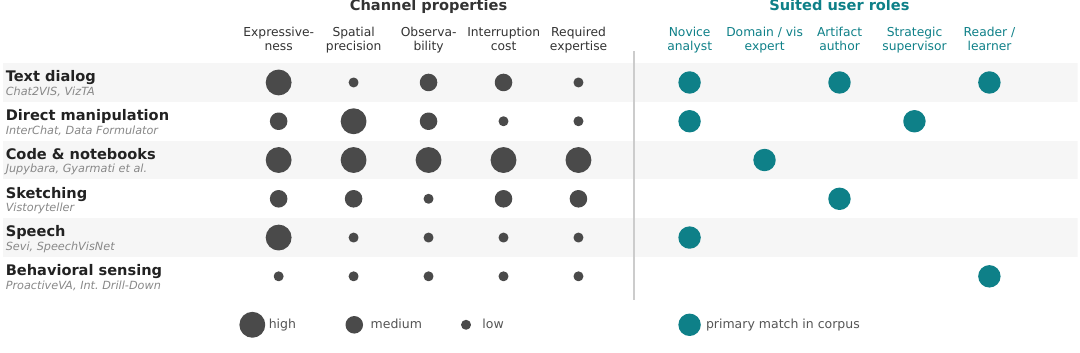}
    \vspace{-2em}
  \caption{Interaction channels observed in the corpus, rated qualitatively on five properties (left block; dot size encodes degree, not desirability) and matched to the user roles they primarily serve in surveyed systems (right block). No channel dominates on all properties, supporting the view of modality selection as a matching problem rather than a search for a single best channel.}
  \Description{A matrix diagram with six rows for interaction channels: text dialog, direct manipulation, code and notebooks, sketching, speech, and behavioral sensing, each with example systems. The left block rates each channel on expressiveness, spatial precision, observability, interruption cost, and required expertise using dot sizes for high, medium, and low. For example, code and notebooks rate high on all five properties, while behavioral sensing rates low throughout. The right block marks the primary user roles each channel serves: text dialog serves novice analysts, artifact authors, and readers; direct manipulation serves novices and strategic supervisors; code and notebooks serve domain and visualization experts; sketching serves artifact authors; speech serves novices; behavioral sensing serves readers and learners.}
  \label{fig:modality-matrix}
  \vspace{-1em}
\end{figure*}

No single interaction modality dominates the corpus, and the evidence argues against universal prescriptions: modalities differ systematically in expressiveness, precision, observability, interruption cost, and the expertise they presuppose (Fig.~\ref{fig:modality-matrix}). Text dialog remains the backbone (51\% of systems support multi-turn dialog) and excels at abstract intent, but is cumbersome for spatial reference. Direct manipulation---lassoing in InterChat~\cite{chen2025InterChatEnhancingGenerative}, drag-and-drop encoding in Data Formulator~\cite{wang2023DataFormulatorAIpowered}---is precise about \emph{where} but weak at \emph{why}. Code and notebook environments (Jupybara~\cite{wang2025JupybaraOperationalizingDesign}, Gyarmati et al.~\cite{gyarmati2025composableagenticautomatedvisual}) maximize observability and surgical control at the price of programming expertise. Richer input channels appear in 63\% of systems: sketching narrative intents in Vistoryteller~\cite{shi2026VistorytellerDesigningDataStories}, speech in Sevi~\cite{tang2022SeviSpeechtoVisualizationNeural} and SpeechVisNet~\cite{zhang2026SpeechtoVisualizationEndtoEndSpeechDriven}. At the passive extreme, behavioral sensing turns the user's actions themselves into input: ProactiveVA~\cite{proactiveVA2026} detects struggle from dwell and toggling patterns, and Intelligent Drill-Down~\cite{zheng2026IntelligentDrillDownLarge} infers intent from interaction trajectories; ProactiveVA can even operate the GUI itself when it takes over a workflow.

These channels serve different human roles, and the corpus covers them unevenly. Novice analysts are served by conversational scaffolding (VizTA~\cite{vizTA2025}, Chat2VIS~\cite{maddigan2023Chat2VISGeneratingData}); domain and visualization experts by inspectable pipelines (PhenoFlow~\cite{kim2024PhenoFlowHumanLLMDriven}, Jupybara~\cite{wang2025JupybaraOperationalizingDesign}); artifact authors by controllable narrative tooling (Vistoryteller~\cite{shi2026VistorytellerDesigningDataStories}, GeoAuthor~\cite{chen2026GeoauthorLinkingTextVisualization}, DataTales~\cite{sultanum2023DATATALESInvestigatingUse}); strategic supervisors by plan-level interfaces (LightVA~\cite{zhao2025LightVALightweightVisual}, FlowForge~\cite{flowforge2026}, CoDA~\cite{chen2025codaagenticsystemscollaborative}); and readers or learners by explanatory overlays (VizTA, ProactiveVA's onboarding tips~\cite{proactiveVA2026}). Designers should treat the modality question as a matching problem over this space (Fig.~\ref{fig:modality-matrix}) rather than a search for a best channel.

\implicationbox{IR-3}{\textbf{Combine language for intent, gesture for scope, and behavior for awareness---and keep proactivity peripheral.} Let users state \emph{what} in language, indicate \emph{where} by direct manipulation, and let behavioral signals inform---not trigger---agent action; proactive help should pass a confidence threshold, appear peripherally, and be ignorable, as in InterChat's interaction linking~\cite{chen2025InterChatEnhancingGenerative}, VizTA's anchored references~\cite{vizTA2025}, and ProactiveVA's ephemeral tooltips~\cite{proactiveVA2026}. \emph{Boundary:} behavioral inference can misread deliberate expert strategies as struggle; monitoring-based interventions need domain calibration, as the bias-monitoring setting of Tytarenko et al.~\cite{tytarenko2026ConfirmationBiasAwarenessVisual} illustrates.}

% Removed standalone guidelines section (P0 revision): guidance content is
% redistributed as situated implications embedded in Sections 3 and 4.
% \input{src/6.guidance}
%!TEX root = ../main.tex

\section{Limitations and Future Research Directions}
\label{sec:future}

% limitations

% While agentic visual analytics represents a significant leap forward, the paradigm is still in its infancy. Our review identifies several critical limitations in current architectures and outlines key research directions to guide the next generation of human-AI collaborative systems.

\subsection{Limitations}

While our survey provides a comprehensive analysis of agentic visual analytics systems, several important limitations exist regarding our scope and methodology.

\textbf{Scope Exclusions.} We prioritize recent LLM-driven architectures, providing limited coverage of early rule-based systems~\cite{shen2023NaturalLanguageInterfaces, zhang2024NaturalLanguageInterfaces}. We also exclude literature focused solely on chart question answering or data storytelling that operates on pre-existing visualizations without active generation. Finally, purely human-centered design studies without novel technical systems are omitted.

\textbf{Methodological Constraints.} The rapid growth of multimodal AI, with most surveyed systems published post-2023, may quickly introduce new paradigms challenging our taxonomy. Additionally, the lack of open-source implementations for many commercial tools (e.g., Tableau's DASH) and certain academic systems constrains our ability to conduct deep, reproducible architectural evaluations.

% \textbf{Scope Exclusions.} First, we provide limited coverage of early rule-based systems, which are extensively documented in previous surveys~\cite{shen2023NaturalLanguageInterfaces, zhang2024NaturalLanguageInterfaces}, to prioritize recent LLM-driven architectures. Second, we exclude literature focused solely on chart question answering (ChartQA) or data storytelling that operates on pre-existing visualizations without generating or revising them. 
% Such systems fall outside our focus on active visual creation workflows, and are only included when serving as assistive components within broader generation pipelines. Finally, we omit purely human-centered design studies that explore cognition and preferences but do not propose novel technical systems.

% \textbf{Methodological Constraints.} 
% Furthermore, the field of agentic visual analytics is experiencing explosive growth. 
% With most surveyed systems published post-2023, the rapid pace of multimodal AI advancements may quickly introduce new paradigms that challenge the completeness of our taxonomy. Additionally, many commercial tools (e.g., Tableau's DASH) and certain academic systems provide limited technical details or lack open-source implementations, which constrain our ability to conduct deep, reproducible architectural evaluations against our framework.

\subsection{Future Directions}

To overcome these limitations, future research must move beyond text-based orchestration and redefine how agents interact with both the analytical environment and the human user.

\stitle{The Agent as a Learner: Continuous Evolution.}
Future systems must develop profile-aware agents that adapt to individual users over time. Fine-tuning agents on real interaction logs allows them to simulate specific personas, requiring longitudinal learning frameworks to measure performance improvements~\cite{misvisfix2026}. At the architectural level, Reinforcement Learning (RL) offers a pathway to true Level 4 autonomy, optimizing the \planner's ability to dynamically restructure workflows. Combined with agent distillation, this self-evolution could significantly reduce the computational overhead of current multi-agent systems.

\stitle{The Agent as a User: GUI-Driven Visual Analytics.}
A forward-looking direction positions the agent as an autonomous digital assistant operating directly within existing GUIs (e.g., Tableau) via accessibility APIs or visual grounding~\cite{proactiveVA2026}. This requires advancing multimodal LLMs to jointly process screen pixels and underlying HTML structures. Furthermore, users must be provided with granular steering capabilities to intervene mid-process~\cite{zhao2025LightVALightweightVisual}. Systems like InterChat~\cite{chen2025InterChatEnhancingGenerative} point toward this future, allowing users to select visual objects to prompt subsequent automated GUI operations.

\stitle{The Agent as a Viewer: Machine-Readable Visualizations.}
As agents assume the \reviewer role, we must rethink visual design to create \textit{Machine-Readable Visual Representations}. Future research should explore generating Semantic SVGs, Knowledge Graph Visualizations, or specialized Agent State Visualizations that facilitate the agent's internal reasoning and self-correction. Designing visual structures optimized for computer vision models rather than human eyes will enhance agent reasoning, promote efficient inter-agent collaboration, and improve the transparency of the visual mapping process.

\section{Conclusion}
\label{sec:conclusion}
Our review of 65 systems reveals that LLMs enable decomposing complex visual analytics tasks into four specialized roles (\planner, \creator, \reviewer, and \contextmanager), driving systems toward increasingly autonomous analytical ecosystems. However, rising AI autonomy does not eliminate human work; it redistributes it from operation toward verification and governance. Our situated, evidence-grounded implications---embedded throughout the workflow and interaction analyses---emphasize reversible delegation, dual-channel review, claim-to-evidence anchoring, and user-controlled memory. Key challenges remain in visual detail blindness, orchestration latency, and human-AI alignment. Addressing these gaps will be critical to realizing trustworthy, democratized agentic visual analytics.

\section*{Ethics and Privacy Statement}

This survey analyzes publicly available, published research on agentic
visual analytics systems; it involves no human participants and no
collection of personal data, and therefore poses minimal direct privacy
risks. However, the technologies we survey carry broader societal
risks: highly autonomous agents can produce hallucinated or misleading
visualizations that users may trust uncritically, and delegating
analysis to agents raises concerns about over-reliance, accountability,
and the deskilling of analysts. Our co-evolutionary framework and
situated design implications directly address this benefit--risk trade-off by
emphasizing human supervision, verifiability, and calibrated trust as
agent autonomy increases.

\bibliographystyle{ACM-Reference-Format}
\bibliography{sample}

\appendix

%!TEX root = ../main.tex

\section*{Supplemental Materials}
\label{sec:supplemental_materials}

All supplemental materials for this paper are available at
\siteurl.
The website includes the project overview, searchable system corpus, and guidance resources that support the analyses and taxonomy presented in this survey.
We provide these materials to improve transparency and reproducibility for future agentic visual analytics research.

\section*{Figure Credits and Copyrights}
\label{sec:figure_credits}

Figures~\ref{fig:example-abc} and ~\ref{fig:example-interface} in our paper include examples adapted from prior work.
All such examples remain under the original copyrights of their respective authors/publishers and are used here only for scholarly comparison and illustration.
The adapted components are credited in the corresponding figure captions and bibliography entries.

\section{Interactive Interface for Corpus and Guidance Exploration}
\label{sec:website_inference}

We provide an interactive companion website at
\siteurl
to support inference, interpretation, and reuse of our survey findings.
The website is designed to help researchers connect concrete systems with our co-evolutionary framework of workflow stages and agentic roles.

\begin{figure*}[tb]
  \centering
  \includegraphics[width=\textwidth]{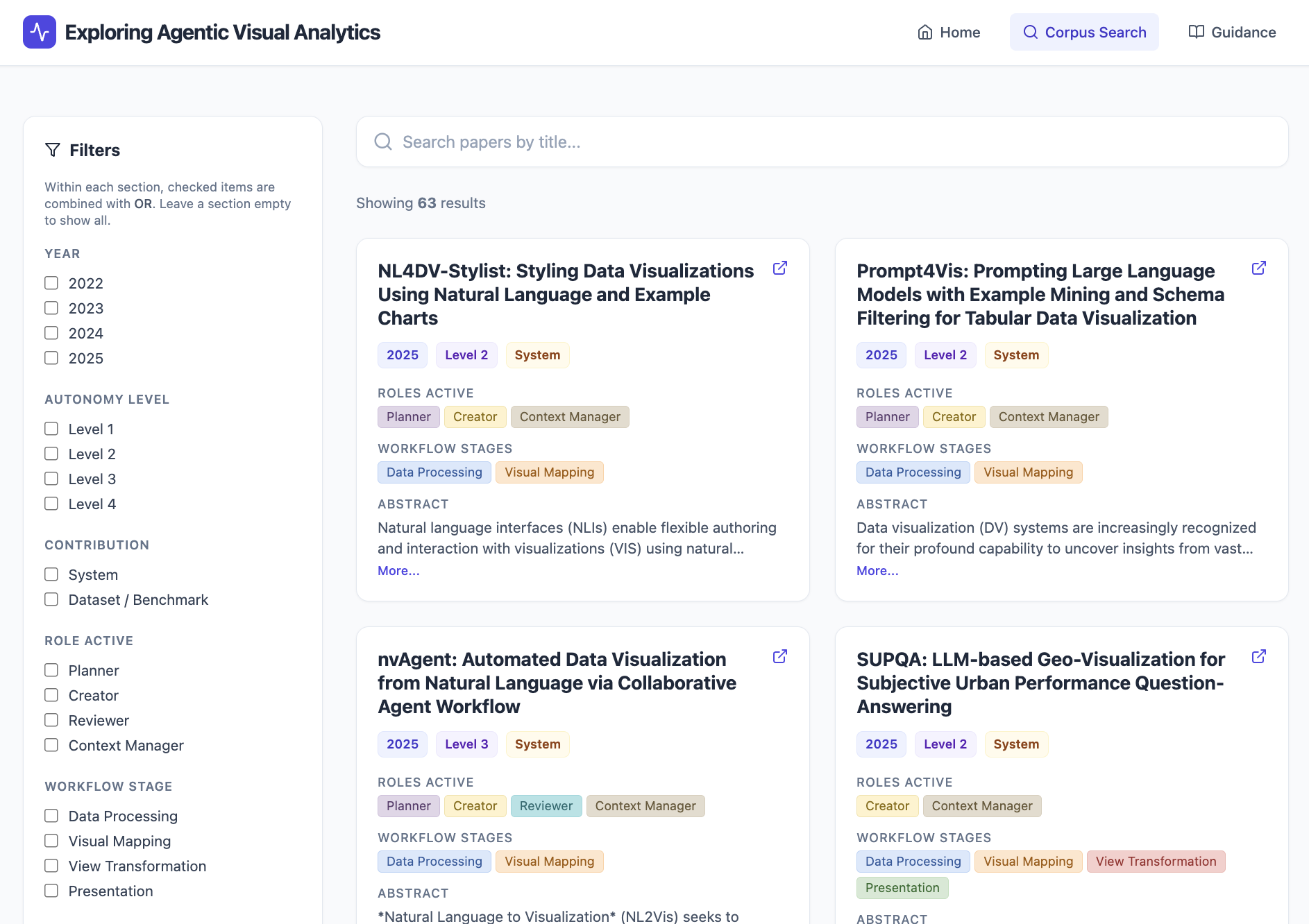}
  \caption{Corpus interface of the interactive companion website. The interface provides multi-dimensional filtering across taxonomy axes and detailed paper cards for evidence-driven exploration.}
  \Description{A screenshot of the companion website's corpus page. A left-hand sidebar offers filters for autonomy level, agentic roles, workflow stages, and publication year; the main area shows a grid of paper cards, each with the system name, metadata, and color-coded tags for roles and workflow stages.}
  \label{fig:appendix-corpus-screenshot}
\end{figure*}

The corpus interface, illustrated in Figure~\ref{fig:appendix-corpus-screenshot}, supports evidence-driven exploration of the surveyed literature.
Users can browse and filter systems across three taxonomic axes: autonomy level, active roles, and workflow stages.
The left-hand panel provides fine-grained filtering options, allowing researchers to isolate systems that meet specific criteria, such as systems published in 2025 that operate at Level 3 autonomy and involve a reviewer role.
The main view displays detailed paper cards that present essential metadata alongside color-coded tags indicating the specific roles and workflow stages instantiated by each system.
This design enables researchers to conduct inferential analysis across the corpus, quickly identifying recurring role combinations, common workflow bottlenecks, and emerging architectural patterns.

\begin{figure*}[tb]
  \centering
  \includegraphics[width=\textwidth]{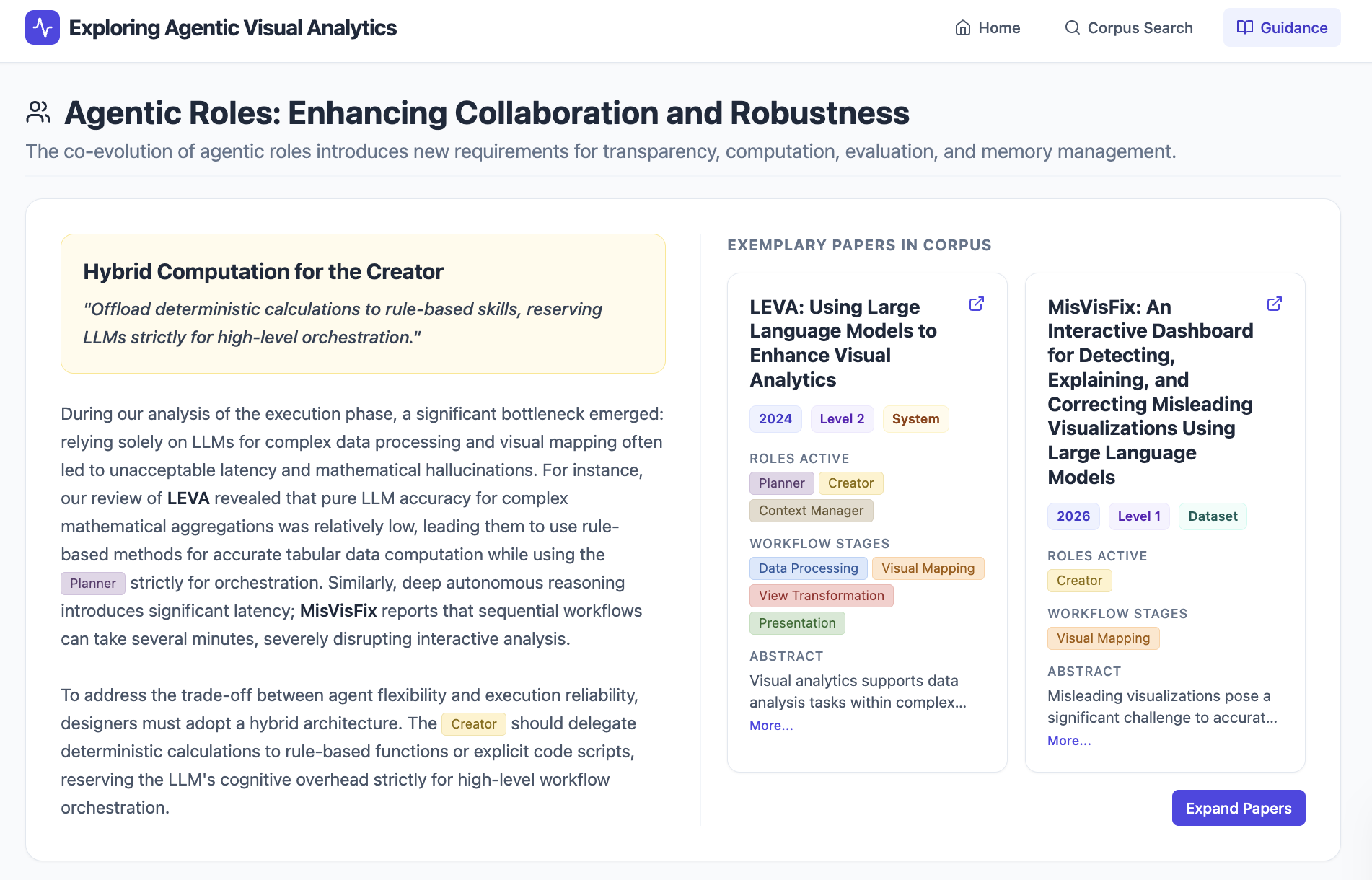}
  \caption{Guidance interface of the interactive companion website. The interface presents actionable design guidelines supported by explicit links to exemplary systems from the corpus.}
  \Description{A screenshot of the companion website's guidance page. Design guidelines are presented as expandable cards, each with a title, explanatory text, and links to exemplary papers from the corpus that instantiate the guideline.}
  \label{fig:appendix-guidance-screenshot}
\end{figure*}

The guidance interface, shown in Figure~\ref{fig:appendix-guidance-screenshot}, translates our synthesis into actionable recommendations.
It bridges the gap between descriptive survey evidence and prescriptive design reasoning.
The interface presents structured design guidelines, such as ``Hybrid Computation for the Creator,'' which addresses the latency and hallucination bottlenecks encountered when relying solely on large language models for complex data processing.
Each guideline is supported by detailed explanatory text and explicitly linked to exemplary papers from the corpus.
For instance, the interface highlights how systems like LEVA and MisVisFix delegate deterministic calculations to rule-based functions, thereby reserving the cognitive overhead of the LLM for high-level workflow orchestration.
Together, the corpus and guidance views form a cohesive interactive tool that empowers researchers to map design goals to appropriate role-workflow configurations and anticipate trade-offs between autonomy and controllability.

\section{Related Work}
\label{sec:related}

Our work builds upon two main research streams: visual analytics systems and natural language interfaces, and agentic AI systems utilizing large language models. We review these areas to position our contributions within the broader research landscape.

\subsection{Visual Analytics and Natural Language Interfaces}

Wu et al.~\cite{wu2022AI4VISSurveyArtificial} systematically review AI techniques applied to visualization through their AI4VIS survey, introducing a WHAT-WHY-HOW taxonomy for visualization data and AI applications. Wang et al.~\cite{wang2022SurveyML4VISApplying} demonstrate ML applications to visualization design and evaluation, revealing rapid growth in automated visualization systems. Hohman et al.~\cite{hohman2019VisualAnalyticsDeep} establish visual analytics' role in addressing AI interpretability and trust challenges using a human-centered framework. However, these foundational surveys predate collaborative agentic systems and focus on individual AI techniques rather than integrated multi-agent workflows.

Zhang et al.~\cite{zhang2024NaturalLanguageInterfaces} present an overview of natural language interfaces for tabular data querying, emphasizing semantic parsing for translating natural language into data manipulation commands. Shen et al.~\cite{shen2023NaturalLanguageInterfaces} demonstrate V-NLI as a complementary input modality for visual analytics, developing seven-stage pipelines covering query interpretation through presentation. Voigt et al.~\cite{voigt2022WhyHowSurvey} analyze two decades of natural language interaction evolution in visualization contexts. However, these approaches focus on separate tools within predefined pipeline architectures without incorporating the agentic reasoning and dynamic generation capabilities of recent LLMs/MLLMs.

Recent LLM-based systems demonstrate practical natural language visualization capabilities. The nvAgent system~\cite{ouyang2025NvAgentAutomatedData} implements three-agent collaborative workflows, achieving significant improvements over baseline approaches.
Talk2Data~\cite{guo2024Talk2DataNaturalLanguage} demonstrates question decomposition for complex analytical queries.
LIDA~\cite{dibia2023LIDAToolAutomatic} implements grammar-agnostic generation with autonomous goal discovery capabilities.
VisEval~\cite{chen2025VisEvalBenchmarkData} provides comprehensive benchmarks for the systematic assessment of LLM-based visualization systems.
While these systems demonstrate effective interaction patterns, they lack comprehensive frameworks for understanding agentic collaboration and systematic workflow orchestration.

\subsection{Agentic AI Systems and Multi-Agent Architectures}

The ReAct framework~\cite{yao2023ReActSynergizingReasoning} pioneered the integration of reasoning traces and task-specific actions in interleaved Thought-Action-Observation loops, making it directly applicable to iterative data exploration.
Guo et al.~\cite{guo2024LargeLanguageModel} present systematic reviews of LLM-based multi-agent systems using five-component frameworks encompassing profile, perception, self-action, mutual interaction, and evolution.
Li et al.~\cite{li2024SurveyLLMbasedMultiagent} categorize LLM agent paradigms into tool use, planning capabilities, and feedback learning using unified LLM-Profiled Roles frameworks.
These foundational frameworks establish core agentic concepts but lack a domain-specific understanding of visual analytics workflows.

Recent implementations demonstrate practical applications for data analysis.
LightVA~\cite{zhao2025LightVALightweightVisual} provides lightweight frameworks supporting task decomposition and insight analysis through recursive planner-executor-controller architectures.
However, these systems operate as monolithic frameworks without the systematic role-based architectures and comprehensive collaboration patterns needed for flexible visual analytics workflows.

While existing work establishes important foundations, significant gaps remain at their intersection. Current visual analytics research focuses on individual tools or simple pipelines without leveraging the collaborative reasoning capabilities of modern agentic systems. Conversely, existing agentic AI research demonstrates powerful coordination but lacks domain-specific understanding of visual analytics workflows and specialized role definitions for analytical collaboration.

Our work uniquely addresses these gaps by providing the first systematic framework for analyzing agentic visual analytics systems through workflow-role decomposition, identifying specific collaboration patterns between specialized analytical agents, and demonstrating how modern LLM capabilities can be orchestrated through multi-agent architectures to support complex, iterative visual analytics workflows. This positioning bridges the gap between agentic AI capabilities and visual analytics requirements, providing both theoretical understanding and practical guidance for this emerging research area.

%% Appendix citations are merged into the main bibliography
%% (ACM-Reference-Format); the separate TVCG-style appendix
%% bibliography (\bibliographyapp) was removed for the CHI submission.

\end{document}